\documentclass[12pt,preprint]{aastex}
\long\def\symbolfootnote[#1]#2{\def\thefootnote{\fnsymbol{footnote}}\footnote[#1]{#2}}

\shorttitle{HST Limits on a Point Source in SN 1987A}
\shortauthors{Graves et al.}

\begin{document}

\title{Limits from the Hubble Space Telescope on a Point Source \\
in SN 1987A}

\author{
Genevieve J. M. Graves\altaffilmark{1,2}, Peter M.
Challis\altaffilmark{2}, Roger A. Chevalier\altaffilmark{3}, Arlin Crotts\altaffilmark{4}, 
Alexei V. Filippenko\altaffilmark{5}, Claes Fransson\altaffilmark{6}, Peter Garnavich\altaffilmark{7}, 
Robert P. Kirshner\altaffilmark{2}, Weidong Li\altaffilmark{5}, Peter Lundqvist\altaffilmark{6}, 
Richard McCray\altaffilmark{8}, Nino Panagia\altaffilmark{9}, Mark M. Phillips\altaffilmark{10}, 
Chun J. S. Pun\altaffilmark{11,12}, Brian P. Schmidt\altaffilmark{13}, George Sonneborn\altaffilmark{11}, 
Nicholas B. Suntzeff\altaffilmark{14}, Lifan Wang\altaffilmark{15}, and J. Craig Wheeler\altaffilmark{16}
}

\altaffiltext{1}{Department of Astronomy, UCO/Lick Observatory, University of California, Santa Cruz, CA 95064 (graves@astro.ucsc.edu)}
\altaffiltext{2}{Harvard-Smithsonian Center for Astrophysics, 60 Garden Street, Cambridge, MA 02138}
\altaffiltext{3}{Department of Astronomy, University of Virginia, P.O. Box 3818, Charlottesville, VA 22903}
\altaffiltext{4}{Institute for Strings, Cosmology, and Astroparticle
Physics, Columbia Astrophysics Laboratory, 550 West 120th Street, Mail
Code 5247, New York, NY 10027}
\altaffiltext{5}{Department of Astronomy, University of California, Berkeley, CA 94720}
\altaffiltext{6}{Stockholm Observatory, AlbaNova, Department of Astronomy, SE-106 91, Stockholm, Sweden}
\altaffiltext{7}{Department of Physics and Astronomy, 225 Nieuwland Science Hall, University of Notre Dame, Notre Dame, IN 46556}
\altaffiltext{8}{JILA, University of Colorado, Campus Box 440, Boulder, CO 80309}
\altaffiltext{9}{Space Telescope Science Institute, 3700 San Martin Drive, Baltimore, MD 21218}
\altaffiltext{10}{Carnegie Institution of Washington, Las Campanas Observatory, Casilla 601, Chile}
\altaffiltext{11}{Laboratory for Astronomy and Space Physics, Code 681, NASA Goddard Space Flight Center, Greenbelt, MD 20771}
\altaffiltext{12}{Current address: Department of Physics, University of Hong Kong, Pokfulam Road, Hong Kong}
\altaffiltext{13}{Mount Stromlo and Siding Spring Observatories, Private Bag, Weston Creek P.O., ACT 2611, Australia}
\altaffiltext{14}{Cerro Tololo Inter-American Observatory, Casilla
603, La Serena, Chile}
\altaffiltext{15}{Institute for Nuclear and Particle Astrophysics, E. O. Lawrence Berkeley National Laboratory, Berkeley, CA 94720}
\altaffiltext{16}{Department of Astronomy, University of Texas, Austin, TX 78712}

\begin{abstract} 
We observed supernova 1987A (SN 1987A) with the {\it Space Telescope
Imaging Spectrograph (STIS)} on the {\it Hubble Space Telescope (HST)} in 1999
September, and again with the {\it Advanced Camera for Surveys (ACS)}
on the {\it HST} in 2003 November.  Our spectral observations cover
ultraviolet (UV) and optical wavelengths from 1140--10266 {\AA}, and
our imaging observations cover UV and optical wavelengths from 2900--9650 {\AA}.
No point source is observed in the remnant.  We obtain a limiting flux of $F_{\rm{opt}}
\le 1.6 \times 10^{-14}$ ergs s$^{-1}$ cm$^{-2}$ in the wavelength range
2900--9650 {\AA}  for any continuum emitter at the
center of the supernova remnant (SNR).  This corresponds to an intrinsic
luminosity of $L_{\rm{opt}} \le 5 \times 10^{33}$ ergs s$^{-1}$.  It is
likely that the SNR contains opaque dust that absorbs UV and optical
emission, resulting in an attenuation of $\sim 35$\% due 
to dust absorption in the SNR.  Correcting for this level of dust absorption would increase our upper 
limit on the luminosity of a continuum source by a factor of 1.54.  Taking into account dust absorption in the remnant, we find a
limit of $L_{\rm{opt}} \le 8 \times 10^{33}$ ergs s$^{-1}$.

We compare this upper bound with empirical evidence from
point sources in other supernova remnants, and with theoretical models
for possible compact sources.  We show that any survivor of a possible
binary system must be no more luminous than an F6 main
sequence star.  Bright young pulsars such as Kes 75 or the Crab
pulsar are excluded by optical and X-ray limits on SN 1987A.  Other
non-plerionic X-ray point sources have luminosities similar to
the limits on a point source in SN 1987A; RCW 103 and Cas A are
slightly brighter than the limits on SN 1987A, while Pup A is slightly
fainter.  Of the young pulsars known to be associated with SNRs, those
with ages $\le 5000$ years are all too bright in X-rays to be
compatible with the limits on SN 1987A.  Examining theoretical models for
accretion onto a compact object, we find that spherical accretion onto
a neutron star is firmly ruled out, and that spherical accretion onto
a black hole is possible only if there is a larger amount of dust
absorption in the remnant than predicted.
In the case of thin-disk accretion, our flux limit
requires a small disk, no larger than $10^{10}$ cm, with an
accretion rate no more than 0.3 times the Eddington accretion rate.  
Possible ways to hide 
a surviving compact object include the removal of all surrounding
material at early times by a photon-driven wind,
a small accretion disk, or very high levels of dust
absorption in the remnant.  It will not be easy to improve
substantially on our optical-UV limit for a point souce in SN 1987A,
though we can hope that better understanding of the thermal infrared
emission will provide a more complete picture of the possible energy
sources at the center of SN 1987A.
\end{abstract}


\section{Introduction} \label{intro}

As the first Local Group supernova
in the era of modern instrumentation, Supernova 1987A (SN 1987A) is at the center of the
investigation into supernova explosions and their aftermath, and into
the formation of compact objects; see review articles by \citet{mcc93} and \citet{arn89}.  
Despite the outpouring of scientific work focused on SN 1987A, the 
initial neutrino burst remains the only evidence for the formation of a compact
object in the supernova event.  No central
point source has yet been detected.  Using the methods described
below, we have obtained an optical upper limit
of $F_{\rm opt} \le 1.6 \times 10^{-14}$ ergs s$^{-1}$ cm$^{-2}$, for a point
source within the supernova remnant (SNR).  At a distance of 51.4 kpc,
assuming 35\% of the emitted flux is absorbed by dust in the remnant,
this corresponds to a luminosity of $L_{\rm opt}\le 8 \times 10^{33}$ ergs
s$^{-1}$.  This is the best 
available limit in this wavelength range.  The best 
upper limits on an X-ray point source are currently the {\it Chandra} limit $L_X \le 5.5 \times 10^{33}$ ergs s$^{-1}$ in the 
2--10 keV band \citep{par04}, and the INTEGRAL upper limit $L_X \le
1.1\times10^{36}$ ergs s$^{-1}$ in the 20--60 keV band \citep{sht04}.

The remnant is located in the Large Magellanic Cloud (LMC), a satellite galaxy of the Milky Way, 
near the massive 30 Doradus \ion{H}{2} region.  The distance to SN 1987A is debated within the range of 50 $\pm$ 5 kpc.  
By comparing the observed angular diameter with models for the emission light curve of the ring, 
\citet{pan99} finds a distance of $51.4 \pm 1.2$ kpc.  \citet{mit02}
use spectroscopic models to fit the expanding atmosphere of the SNR, and obtain a distance modulus of $18.5 \pm 0.2$, which corresponds to $50.1
\pm 4.0$ kpc, and is consistent with the result of \citet{pan99}.  In 
this paper, we use a distance of $51.4$ kpc.  

SN 1987A is the only supernova for which we have 
detailed observations of the region predating the supernova outburst.  Based on the coordinates 
of the supernova, the B3I blue supergiant, Sanduleak -69$^{\circ}202$ 
\citep{san69} was identified as a likely progenitor (e.g.,
\citealt{kir87}; \citealt{wal87}; \citealt{wes87}; \citealt{whi87}).  SN 1987A 
was a Type II supernova (SN II), as 
determined by the strong hydrogen lines in its spectrum, but unlike
normal SNe II, 
it did not reach its maximum luminosity until $t \approx 80$ days.  Moreover, its maximum 
luminosity was only $1/10$ the typical maximum for SNe II.  Subsequent modeling of 
this peculiar SN II (see e.g. Woosley 1988; Nomoto, Shigeyama, \&
Hashimoto 1987) matched the evidence with a progenitor mass of $\sim 20 M_{\odot}$, 
corroborating the identification of the blue supergiant Sanduleak -69$^{\circ}202$ as the progenitor.  This accounts for the 
differences between SN 1987A and normal SNe II, which
generally result from a core collapse inside a red supergiant.

One striking feature of SN 1987A is the set of ring-like structures that surround the central 
remnant.  These rings are material ejected from the aging star that
was ionized by ultraviolet (UV) 
photons from the supernova explosion (see Lundqvist \& Fransson 1996).
The presence of rings in the SNR strongly suggests that SN 1987A is an
aspherical system.  This is corroborated by the polarization of light
from the remnant, the shape of the ejecta, the
kinematics of the debris, the ``mystery spot,'' and an epoch of
asymmetry in the emission line profiles known as the ``Bochum event.''
These all indicate an aspherical shape for the supernova \citep{wan02}.  
The central ring of SN 1987A appears 
in {\it Hubble Space Telescope (HST)} images as a $1.7''$ by $1.2''$ ellipse, interpreted as a circular ring lying in the equatorial plane of the progenitor star \citep{cro91}.  
\citet{pan91} give an inclination angle of 42.8$^{\circ} \pm 2.6^{\circ}$, derived from the UV 
light curves of the ring.  

The progenitor star, Sanduleak -69$^{\circ}202$, is estimated to have
had a zero-age main-sequence (ZAMS) mass $M \sim$ 20
$M_{\odot}$.  This mass lies in the range where it is uncertain
whether core collapse will form a neutron star or a black hole.
Simulations by \citet{fryer99} show that both $15 M_{\odot}$ and $25
M_{\odot}$ stars will collapse initially to a neutron star, but that the
neutron star left by the collapse of the $25 M_{\odot}$ star will
accrete gravitationally bound material and further collapse to a black
hole. The mass boundary between these possible end states depends on
the neutron star equation of state, the physics used in the model, and other factors.  The
asymmetry of the explosion also may have influenced the final
outcome.  With a progenitor mass of $M \sim 20 M_{\odot}$, the compact remnant of SN 1987A could be either a neutron
star or a black hole.

To date, there is no direct evidence for a compact object at
the center of the SN 1987A remnant.  \citet{sun92} give an upper limit on the energy contribution from a central continuum source of $8 \times 10^{36}$ ergs s$^{-1}$ for 1500 days after outburst. 
The current spectrum of SN 1987A is dominated by emission
lines from gas in the remnant, powered by the radioactive decay of $^{44}$Ti; see \citet{koz98} for a detailed discussion of the line emission.  Radio astronomers have
been looking for a pulsar in the remnant of SN 1987A, but no pulsar has been seen.  \citet{gae97} include a summary of the negative
results of this search.  A claim was made for a submillisecond optical pulsar in SN 1987A \citep{kri89}, but it was 
later retracted \citep{kri91}, the false signal being attributed to electrical noise in the data acquisition system.
More recently, \citet{mid00} have presented evidence for a millisecond pulsar, but their results have not been confirmed by
others.  \citet{fry99} point out that even if such a pulsar
were beamed away from our line
of sight, we would expect to see a significant luminosity contribution from the surrounding debris, acting as a
calorimeter.

In this paper, we place an upper limit on the luminosity from a possible continuum source in 
SN 1987A that is three orders of magnitude lower than previous
values.  We do this using recent observations of the SNR which
take advantage of the high sensitivity and powerful resolution of the {\it
Advanced Camera for Surveys (ACS)} on {\it HST} to probe the remnant deeply in five filters, ranging from near-UV
to near-infrared (IR) wavelengths.  Our observations and data reduction are summarized in \S\ref{obs}.  
We present our upper limits from the spectrum and the images in \S\ref{limits}, and discuss the possible
effects of internal dust absorption.  Section \ref{sources} considers our upper limits in the case of a potential binary companion or an optical pulsar, and compares the limits on a compact 
object in SN 1987A with point sources in other SNRs.  Various accretion 
models are summarized in \S\ref{acc} and are discussed in the context of our upper limits on a 
continuum source.  Fallback models for SNRs are described in \S \ref{fall}, and \S \ref{conc} summarizes our
conclusions.   

\section{Observations} \label{obs}

\subsection{{\it STIS} Observations} \label{stisobs}

During the month of 1999 September, the Supernova INtensive Study
(SINS) group made a series of spectral observations of
SN 1987A using the {\it Space Telescope Imaging Spectrograph (STIS)}
on board the {\it HST}.  These spectra were reduced using the on-the-fly calibration system provided by the Space Telescope
Science Institute.  The observations covered the entire wavelength range
available from {\it STIS}, 1140--10266 {\AA} (Table
\ref{stis}).  The {\it STIS} observations
were as follows: in the far-UV with the $0.5''$ slit using the G140L grating; in the near-UV with the $0.5''$
slit and the G230L grating; in the optical with the $0.2''$ slit and the G430L grating; and in the optical
through near-IR with the $0.2''$ slit and the G750L grating.  The slit was oriented to
cross the central region of the supernova, and to exclude two bright foreground stars located
$\sim1.3''$ away on the sky (see Figure \ref{slitpos}).  Each observation has multiple exposures and spatial dithers to remove the cosmic rays,
flat-field features, and hot pixels.  Additional cosmic ray cleaning was done on each individual exposure.  The
spectra were averaged over these multiple exposures, using a weighted median algorithm.  

We used the IRAF\symbolfootnote[1]{IRAF is distributed by the National Optical Astronomy Observatories, which are operated by the Association of Universities for Research in Astronomy, Inc., under cooperative agreement with the National Science Foundation.} program {\it apall} to collapse
the two-dimensional (2-D) spectra into one-dimensional (1-D) spectra for each grating.  A compact remnant or accretion system in the SNR 
would appear as a point source at the center of the supernova debris,
and as a continuum source along the center of the 2-D spectrum.
We extracted the 1-D spectrum from the geometric center of the debris
region, with a width of 4 pixels.  This is the width of the point
spread function (PSF) in the 2-D spectra.  The flux was summed over this 4-pixel width to include the total flux from the
central object.  The IRAF program {\it scombine} was used to
combine the spectra from the four gratings into one spectrum spanning
the wavelength range 1140--10266 {\AA}.  A steep rise is
observed in the spectrum over wavelengths shorter than 1400 {\AA}.
This is probably due to noise in the far-UV data multiplied by a large
flux-correction factor.

Interstellar reddening is significant in the direction of SN 1987A,
so to place limits on the intrinsic properties of a source the spectrum must be corrected for this effect.  The reddening in
the 30 Doradus region is highly variable; \citet{pan00} have
calculated the reddening for over 2,000 individual stars within 30 pc
of SN 1987A, and find a large spread of reddening values, with an
average value of $E(B - V)$ = 0.203 $\pm 0.072$ mag.  

The best value
for the reddening in the direction of SN 1987A should be the value for
stars in the immediate vicinity of the SNR.  We use the reddening
derived by \citet{scu96} for Star 2, one of the bright stars directly
adjacent to the SNR.  \citet{scu96} find a total of $E(B - V) = 0.19
\pm 0.02$ mag of reddening in the direction of Star 2.  They show that
the observed reddening of Star 2 is well fit by a two-component
reddening model, in which the Galactic reddening curve of \citet{sav79}
and the \citet{fit86} reddening curve for the 30 Doradus region
are combined in a ratio of 1:2.  The extinction function for the 30 Doradus region derived by \citet{fit86} differs significantly from that of the
Galaxy in the UV: the bump at 2175 {\AA} is significantly weaker and flattened compared to
Galactic absorption, there is greater absorption at wavelengths shorter than 2000 {\AA}, and from there, the
extinction rises steeply toward the far-UV.  It also differs from the
standard LMC extinction function.  Following \citet{scu96}, we use the
\citet{sav79} reddening curve with $E(B - V)_{\rm Galactic} = 0.06$ mag
and $R_V = 3.2$ for the Galactic reddening, and the \citet{fit86}
reddening curve with $E(B - V)_{\rm LMC} = 0.13$ mag and $R_V = 3.1$ for
the LMC reddening.

\subsection{{\it ACS} Observations} \label{acsobs}

The images were taken in 2003 November by SINS, using the
{\it High Resolution Camera (HRC)} and {\it ACS} on board {\it HST}.
We obtained images in five filters: F330W,
F435W, F555W, F625W, and F814W, which correspond to the {\it HRC} UV, Johnson B, Johnson V, Sloan Digital Sky
Survey (SDSS) r, and broad I bands respectively.  These five filter
bands were chosen to cover the entire wavelength range from 2900--9650
{\AA}, spanning the near-UV to the near-IR.  

The images were drizzled to combine dithered exposures and remove
cosmic rays (see \citealt{fru02}).  These observations are summarized in Table
\ref{acsdata}, and the imaging data are shown in Figure \ref{images}.
A visual inspection of the images shows that no point source is apparent in
the SNR.  A quantitative upper limit based on these
images will be presented in \S \ref{image}.

\section{Upper Limits on the Brightness of SN 1987A} \label{limits}

\subsection{Upper Limits from the Spectrum} \label{spec}

The spectrum of the central source in SN 1987A corrected for
interstellar extinction is shown in
Figure \ref{final_cont_image}.  To get a limit on a point source,
we looked between the emission lines for the spectrum of an underlying
continuum object. The spectrum is dominated by H$\alpha$, Mg, Fe, O, Ca, and Na emission features in the cool gas surrounding the compact remnant ($T \approx$
130--160 K).  There are also a variety of metal lines, powered by the reprocessing of radioactive decay from
$^{44}$Ti; see \citet{chu97} and \citet{wan96} for a thorough discussion of the emission spectrum of SN 1987A.  Underneath this
emission spectrum, there is a continuum flux density, on the order of
$F_{\lambda} = 10^{-17}$ ergs s$^{-1}$
cm$^{-2}$ {\AA}$^{-1}$, which corresponds to $L_{\lambda} =
3\times10^{30}$ ergs s$^{-1}$ {\AA}$^{-1}$ at a distance of 51.4 kpc.  This continuum flux could include flux from weak emission lines, continuum radiation from the gas in the SNR, continuum emission from a compact object in the center of the remnant, or very likely a combination of emission from all of these sources.  Our measurement of this continuum flux from SN 1987A places an upper limit on the contribution from a compact object.  

In our continuum fit, we used only the wavelength ranges that lie {\it between}
conspicuous emission lines: 1920--2070 {\AA}, 3875--4020 {\AA}, 4615--4665 {\AA}, 5405--5575 {\AA}, 6010--6210 {\AA}, 6780--7080 {\AA}, 7555--8060 {\AA}, 8925--9125 {\AA},
9400--9620 {\AA}, 9830--9980 {\AA}, and 10100--10150 {\AA}.

We fit a polynomial function to these regions of the spectrum.  The
continuum fit is plotted with the spectrum in Figure
\ref{final_cont_image} as the green dashed line.
We found that a seventh-order Chebyshev polynomial gave the best fit.
We then integrated this function across the spectrum to obtain an
upper limit on the total luminosity of the central source.  In the
optical range covered by our imaging data (2900--9650 {\AA}) the
integrated continuum flux was found to be $5.9 \times 10^{-14}$ ergs
s$^{-1}$ cm$^{-2}$.  At 51.4 kpc, this corresponds to an upper limit
on the optical luminosity of $L_{\rm opt} \le 1.9 \times 10^{34}$ ergs
s$^{-1}$, roughly five times the optical luminosity of the Sun.  In
the UV range (1400--2900 {\AA}) the integrated continuum flux was $5.5
\times 10^{-14}$ ergs s$^{-1}$ cm$^{-2}$, therefore $L_{\rm UV} \le 1.7 \times 10^{34}$ ergs s$^{-1}$.

\subsection{Upper Limits from Images} \label{image}

We examined {\it ACS} images of SN 1987A taken in 2003 November
to determine an upper limit on a point source in the SNR.  We
approached this task by asking how bright a point source in the center
of the debris could elude detection. We determined the
geometric center of the remnant in several different ways: by fitting
an ellipse to the inner ring of the SNR, by fitting a circular
aperture around the central debris, and by eye.  These measurements
all agreed within two pixels.  For observations made with {\it HRC},
the pixel size varies across the detector from $\sim0''.0275$ to
$0''.0287$ on a side \citep{mac03}.  Using a pixel size of $0''.028$
per pixel, a two pixel uncertainty in the centerpoint of the remnant
corresponds to $0''.056$.  Even if the supernova explosion
imparted a kick velocity of 1000 km s$^{-1}$ to the compact object, it
would only have travelled a maximum of $0''.07$ on the sky by the time
of our observations, or 2.5 pixels.  Allowing for a high initial kick
velocity and a two-pixel uncertainty in the centerpoint of the
remnant, we can draw a circle in the center of the remnant with a
radius of 4.5 pixels ($0''.126$) that should contain any compact object
resulting from the supernova event.  This 4.5 pixel ring is shown as
the red ring in
Figure \ref{images}f. 

If a point source exists within the remnant, it must be too dim to be
detected by visual inspection.  This depends both on the flux from the
source and how it hits the detector; if most of the light from the
point source falls on one pixel in the detector, it will be more
easily visible against the glowing debris in the center of the remnant
than if the point source falls between pixels and its light is
distributed over as many as four pixels.  The point spread functions
(PSFs) of actual sources in each image should give a reasonable
sampling of this effect. To determine the brightest point
source that could hide within the remnant, we used the software
package {\it DAOPHOT} \citep{ste87} to do photometry on the entire
field in each image, producing a photometry catalog of around 1,800
sources for each imaging band.  We then inserted actual PSFs of field
stars with a 3-pixel radius into the central debris of SN 1987A, at the centerpoint shown by
the red dot in Figure \ref{images}f.  The insertion was done using a
pixel-by-pixel substitution of the PSF into the central debris.  For
each pixel within a 3-pixel radius of the central point, the count
level of the PSF pixel was compared with the count level of the
corresponding pixel in the central debris, and the brighter pixel was placed
in the artificial image.  This has the effect of inserting the full
radius of the PSF
into the central debris, without creating a dark ring around the
central PSF.

Starting with bright sources that were clearly visible, we inserted
progressively fainter objects until we were unable to detect the
inserted source, repeating the process for each imaging band.  Figure
\ref{f435_fakes} shows this process for the B-band filter
(F435W).  The {\it DAOPHOT} catalog gave the total counts for the
limiting objects, measured in a 4-pixel aperture.  We used aperture
corrections derived from white dwarfs in the image to correct to an infinite aperture,
giving us the total counts of the brightest point source that could remain
undetected in SN 1987A.  

To test the robustness of this result against
the uncertainty in the centerpoint and the movement of a compact
remnant due to an initial kick velocity, we repeated the measurement
in each filter at four other points within the 4.5 pixel ring shown in
Figure \ref{images}f.  In four of the five filters, the measured upper
limits were within 5\% of the original results for all measured
locations in the 4.5 pixel ring.  In filter F555W, the measured upper
limit varied by as much as 14\% between the various measured
locations.  In \S \ref{dust}, we will show that dust absorption within
the remnant is likely to raise our upper limit by a factor of 1.54 (54\%), thus
the uncertainties due to dust absorption are
several times larger than the uncertainties due to the centerpoint determination and
the movement of the remnant due to an initial kick velocity.

To convert the measured counts to physical flux units, we assumed a flat spectrum across each
imaging band.  We then used the IRAF package {\it
synphot}\symbolfootnote[1]{{\it Synphot} is a part of the software product STSDAS.  STSDAS is a product of the Space Telescope Science
Institute, which is operated by AURA for NASA.} 
to convert a count level into a constant flux per angstrom
($F_{\lambda}$) for each filter.  

In detail, we did this using the {\it synphot} task {\it calcphot} to
compute the total counts produced by an input spectrum.  The spectrum
was specified as a flat spectrum with constant $F_{\lambda}$.  At this
stage, we also included a correction for the reddening due
to dust in the LMC and our Galaxy.  As in section \ref{stisobs}, we used
an $E(B - V)_{\rm Galactic} = 0.06$ mag with $R_V = 3.2$ for
the Galactic foreground reddening and $E(B - V)_{\rm LMC} = 0.13$ mag with
$R_V = 3.1$ for
the reddening contribution from the LMC.  We used the \citet{sav79}
reddening curve for the Galactic
reddening, and the \citet{how83} LMC reddening curve provided in
the {\it synphot} task {\it calcspec}.  In this case, the difference in the
reddening between the Howarth LMC curve and the Fitzpatrick curve specific to the 30
Doradus region is irrelevant, because it does not affect the
wavelength range covered by our imaging (2900--9650 {\AA}).  

The flat
spectrum, corrected for reddening, was used as the input spectrum for
{\it calcphot}, and the constant flux level $F_{\lambda}$ was varied
until the output count level matched our upper limit on total count
level for a point source in each filter.  The
upper limits for each filter are given in Table \ref{filterlimits}.  The value of $F_{\lambda}$ for each filter was used over the entire
filter width, as defined by the ``applied filter width'' in Table \ref{filterlimits}.  The applied filter width for each filter was chosen so
that there were no gaps between filters.  The transition wavelengths
between adjacent filters were determined by the wavelengths at which
the {\it ACS/HRC} filter throughput curves crossed. This means that at all
wavelengths in the 2900--9560{\AA} range, the corresponding value of
$F_{\lambda}$ is that belonging to the filter with the highest
throughput value for that wavelength.  For comparison, the upper limit
on $F_{\lambda}$ for each filter is plotted against the 1999 {\it STIS}
spectrum in
Figure \ref{final_cont_image} as a thick solid line, whose color
corresponds to the color of the filter (violet for UV-band, blue for
B-band, yellow for V-band, red for R-band, and maroon for I-band).
All of the upper limits from the imaging data lie below the continuum
fit to the spectrum, so the 2003 November imaging data
give a lower limit than the 1999 September spectroscopic data.  The
extracted spectrum includes more background than the {\it ACS} images,
which use the full resolution of the {\it HST}.

To obtain a total optical luminosity limit, the
flux density per Angstrom $F_{\lambda}$ for each filter was integrated across
the applied filter width, then the contributions from each filter were
summed.  The summed flux was converted to an upper limit on the total
optical flux from a point source in the remnant of SN 1987A:
$F_{\rm opt} \le 1.6 \times 10^{-14}$ ergs s$^{-1}$ cm$^{-2}$.  At a distance
of 51.4 kpc, this corresponds to an intrinsic luminosity of $L_{\rm opt} \le 5 \times 10^{33}$ ergs s$^{-1}$.  This is comparable
to the optical luminosity of the Sun, and is three orders of
magnitude lower than any previously published upper limit for this source.

\subsection{The Effects of Dust Absorption} \label{dust}

By about 650 days after the first detection of SN 1987A, most of the bolometric luminosity of 
the SN was being emitted in the IR.  This has been interpreted as radiation from dust that
condensed within the SNR.  Dust began to form as early as day 100 and the formation of new dust 
appears to have dropped dramatically after day 650 \citep{bou93}.  There is no evidence for spectral 
reddening due to the dust in the SNR, which suggests that the dust is
opaque ``black dust'' made of particles that are large compared to the
wavelength of these observations.

An analysis of data across the IR bands by \citet{bou96} shows that 97{\%} of the 
bolometric luminosity of SN 1987A was being emitted in the IR at day 2172.  Since that energy is initially 
emitted at higher wavelengths, dust absorption has a significant effect on the UV and optical 
luminosity of the SNR.  To determine the effects of dust on our upper limit, we will assume 
that 97{\%} of the UV and optical luminosity of SN 1987A was being absorbed by dust at day 2172.
A substantial portion of the observed IR luminosity may be due to collisional heating, rather than 
the re-absorption of UV or optical radiation, so the attenuation may be less than 97{\%}.  

An attenuation of 97{\%} at day 2172 corresponds to an effective
optical depth of $\tau_{\rm eff} =3.5$.  
This is equivalent to the mean optical depth, if the dust is distributed uniformly.  However, 
there is evidence that some significant portion of the dust is in clumps \citep{bou93,bou96}.  Using the extinction model by \citet{nat84} for dust clumps in
a Poisson distribution, 
\begin{equation}
\tau_{\rm{eff}} = N(1-e^{-\tau_c}), 
\end{equation}
where $N$ is the
number of absorbing clumps along the line of sight and $\tau_c$ is the
optical depth of the individual clumps.  In the limit of opaque dust
clumps, $\tau_c \rightarrow \infty$ and $\tau_{\rm{eff}} \rightarrow
N$. 

With dust formation dropping off precipitously after day 650, it seems
reasonable to assume that dust has neither been formed nor destroyed since
Bouchet's observations on day 2172.  In this 
case, the optical depth should scale as the column density of the expanding dust cloud, that 
is as $1/t^{2}$ for homologous expansion.  Scaling from day 2172 to
our imaging data at about day 6110, 
we have an optical depth of $\tau_{\rm eff} = 0.44$ for uniformly
distributed dust.  In the clumped dust scenario, with $\tau_{\rm eff}
\rightarrow N$, the optical depth also scales as the column
density, and both configurations of dust in the envelope give the same
value for the optical depth, $\tau_{\rm eff} = 0.44$.
This value of $\tau_{\rm eff}$ corresponds to an 
attenuation of $\sim 35${\%}, so dust in the remnant may be absorbing
35\% of the UV and optical luminosity of SN 1987A.
The actual UV and optical luminosity of the central source is likely
to be 1.54 times the limit we have measured, if the dust is uniform, or clumped on small scales.

This argument for the effective optical depth for clumped dust only applies if the dust is clumped 
on scales that are smaller than the size of the central source that it obscures.  In the worst case 
scenario, there would be a single, large clump of dust directly along the line of sight to the central part 
of the SNR.  In this case, the expansion of the SNR would not thin out the dust, and the clump would have the 
same size and optical depth that it had at day 2172, absorbing 97{\%} of the UV and optical flux from the central 
source.  Again, this is based on the assumption that the observed IR flux at day 2172 was entirely due to the 
re-absorption of UV and optical radiation from the central source, neglecting collisional heating.  This 97{\%} 
attenuation corresponds to an upper limit on the optical and UV flux
from SN 1987A that is $\sim 30$ times the 
limit we have measured.  While this worst case scenario for dust absorption must be considered a possibility, 
it should not prevent us from exploring the physical implications of
our upper limit in the more likely case of absorption by uniform dust,
or dust clumped on small scales.  Throughout the remainder of this paper, we will use
an upper limit corrected
for absorption at the level of 35\%, as in the uniform or clumped dust
scenarios described above.  This results in an optical upper limit
$L_{\rm{opt}} \le 1.54 \times (5 \times 10^{33})$ ergs s$^{-1}$, or
$L_{\rm opt} \le 8\times10^{33}$ ergs s$^{-1}$.

\subsection{Can Infrared Observations Help?}

\citet{bou03} observed SN 1987A with the Thermal-Region
Camera and Spectrograph (T-ReCS) on Gemini South and resolved a point
source at 10$\mu$m.  This source corresponds to the central ejecta of
the remnant, inside the inner ring.  According to \citet{fra02}, 6000
days after outburst, the radioactive decay of $^{44}$Ti should still be
injecting enough energy into the SN ejecta to power a luminosity of
10$^{36}$ ergs s$^{-1}$.  The energy of this decay is enough to power
the mid-IR flux observed by \citet{bou03}.  Assuming 10$^{36}$ ergs
s$^{-1}$ of thermal radiation, they constrain the temperature of the
dust in the ejecta to $90 \le T \le 100$ K.  No energy source other
than $^{44}$Ti is needed to account for the mid-IR flux of the ejecta.
Spectroscopic observations of the ejecta with the Infrared Spectrograph on the {\it Spitzer Space
  Telescope} might be able to determine the energy mechanism producing
the observed mid-IR emission of the central ejecta, and either confirm
$^{44}$Ti decay as the energy source, or reveal the need for another
contributing energy source.

\section{Observed Limits Applied to Possible Continuum Sources} \label{sources}

\subsection{Possible Binary Companion} \label{binary}

A number of evolutionary models have been proposed for SN 1987A that model the progenitor 
star as a member of a binary system.  The earliest of these models were those of 
\citet{jos88}, \citet{fab88}, and \citet{pod89}, followed by \citet{loo92}, and others.  A more recent simulation by 
\citet{col99} models the circumstellar nebula and ring structure of SN 1987A by supposing a 
merger between the progenitor of SN 1987A and a binary companion before the supernova explosion. 

If a companion star remains in the SNR, it would have to fit below our upper 
limit to escape detection.  We have an upper limit $L_{\rm opt} \le 8
\times 10^{33}$ ergs s$^{-1}$, or $L_{\rm opt} \le 2 L_{\odot}$.  This
restricts a surviving main-sequence companion to stars of type F6 or later,
if there is clumped dust in the remnant.  Our limit would also, of
course, be consistent with the presence of a white dwarf companion.

\subsection{Comparison with Other Point Sources in SNRs} \label{snrs}

It is instructive to compare X-ray and optical limits on SN 1987A with point sources
found in other SNRs.  \Citet{cha01} identify four types of X-ray point
sources that may be associated with SNRs.  These are classical
pulsars, anomalous X-ray pulsars (AXPs), soft gamma repeaters (SGRs),
and a catch-all category for non-plerionic X-ray point sources that do
not appear to fit in any of the previous three categories.  Of the
classical pulsars, the ones of most interest with respect to SN 1987A
are the young pulsars.  Table \ref{compare} shows X-ray and optical
luminosities for a number of other point sources detected in SNRs.
The table combines data taken from Table 3 of \citet{cha01} and Table
2 of \citet{zav04b}, as well as some additional optical upper limits.
The table includes only point sources that appear to
be associated with SNRs.  

From Table \ref{compare}, we can see that the X-ray and optical upper
limits on the luminosity of a central source in SN 1987A are low
enough to present an interesting comparison with other point sources
in SNRs.  Of the young pulsars listed in Table \ref{compare}, only
those with ages $> 1 \times 10^4$ yr are consistent with the limits
for SN 1987A.  Standard models for pulsars give a declining pulsar
power with age, making the case for a pulsar in SN 1987A problematic.
The youngest of the young pulsars are all too bright in
X-rays to fit within the X-ray limits.  In fact, all of the sources listed
in Table \ref{compare} which are consistent with the limits on SN
1987A have ages $\ge 5000$ yr, with the exception of Pup A; a source
such as Pup A could remain hidden in the central debris of SN 1987A.
It should be noted however that the age given in the table for Pup A, 3000 years, is derived
from the kinematics of fast moving oxygen filaments in the remnant.
Ages derived from the X-ray temperature are 5000-10,000 yr (see
\citealt{win85}), so there is significant uncertainty in this result.
Overall, this comparison shows that any point source in the remnant of
SN 1987A would have to be fainter than the X-ray point sources
detected in other very young supernova remnants.

\subsection{A Pulsar in SN 1987A?} \label{pulsar}

Since its outburst, astronomers have been looking for a pulsar in the remnant of SN 
1987A as confirmation of the widely-accepted theory that pulsars are born in supernovae.  Despite repeated
observations, no radio pulsar has been detected to date.  The paper by
\citet{gae97} includes a summary of the negative
results of this search.  The observations in
\citet{gae97}, taken with the Australia Telescope Compact Array (ATCA), show an asymmetric shell of
radiation around the stellar remnant, but no central point source.  The lack of a central radio source and
radio pulsations is evidence against the presence of a radio pulsar.

\citet{mid00} report optical timing observations that may imply the existence of an
optical pulsar in the remnant of SN 1987A with a period of 2.14 ms.  Their data were taken between 1992 and
1996 on multiple telescopes, including those at the {\it European
  Southern Observatory (ESO)} and the {\it
HST}. The
signal is detected sporadically on all of the instruments, with a signal detected in 21 out of the 78
nights of observing.

To escape detection in optical images of SN 1987A, a potential pulsar
would have to have either a small dipole magnetic field, or a very
large one.  \citet{oge04}
use a previous upper limit on the bolometric luminosity of SN 1987A to
constrain the properties of a pulsar in the remnant.  They show that
an upper limit on the luminosity constrains the initial rotation period $P_0$ and the
magnetic dipole moment $\mu$ such that there is an elliptical region
of $P_0$-$\mu$ space that is excluded as possible values for a pulsar
in SN 1987A.  Using data from \citet{sod99}, they derive a limit
$L_{\rm bol} \le 3 \times 10^{34}$ ergs s$^{-1}$ at $t=11.9$ years.  
The corresponding ellipse in $P_0$-$\mu$ space excludes the range of
``normal'' magnetic dipole moments for the observed range of typical
pulsar periods.  This means that a young pulsar in the remnant of SN
1987A must have either a weak magnetic dipole moment or a very strong
magnetic dipole moment, pushing it into the range of a magnetar.  For
a pulsar with an initial period of 0.03 seconds ($\sim$ the Crab
pulsar), \citet{oge04} find that the magnetic field is limited to $\mu < 2.5 \times 10^{28}$
Gauss cm$^3$, or $\mu > 2.4 \times 10^{34}$ Gauss cm$^3$.  For the
pulsar claimed by \citet{mid00} with a period of 2.14 ms, they find the magnetic
field is $\mu < 1.1\times10^{26}$ Gauss cm$^3$ or $\mu > 2.4 \times
10^{34}$ Gauss cm$^3$.  These values are outside the bounds of normal
pulsars.  For comparison, conventional pulsars have
magnetic moments of $\sim 10^{30}$ Gauss cm$^3$.  

If a pulsar in SN 1987A were shrouded in thick dust, the dust
would act as a calorimeter, reradiating energy beamed into it by the
pulsar.  This would appear as IR flux from the central ejecta.
\citet{bou03} observe a mid-IR luminosity of $\approx 10^{36}$ ergs
s$^{-1}$.  This is
consistent with the IR luminosity assumed by \citet{oge04}, therefore
their conclusions would apply even in
the case where dust is obscuring more than 97\% of the pulsar luminosity and reradiating the energy
in the infrared.  

Based on the arguments presented here, and also upon the comparison
with other young pulsars, it seems unlikely that the remnant of SN
1987A currently harbors a pulsar.  Theoretical models of pulsar formation
(ie. \citealt{bla83}) give formation times for pulsars of up to $10^5$
yr, so it may be that a pulsar has not yet had time to form in the
remnant.  

\section{Accretion Models} \label{acc}

\subsection{Overview of Accretion} \label{accover}

In the aftermath of a supernova, models predict a certain amount of
fallback, in which gravitationally bound debris from the explosion
eventually accretes onto the compact object at the center. For SN
1987A, the amount of matter that accretes onto the central object
during fallback was
estimated by \citet{che89} to be $M_{\rm acc} \approx 0.15 M_{\odot}$.
He assumes spherical accretion, in which the amount of fallback is
strongly dependent upon the sound speed $c_s$ of the gas in the remnant;
the accretion rate $\dot{M}_{\rm{acc}}$ scales as $\dot{M}_{\rm{acc}} \propto c_s^{-3}$
\citep{che89}.  
More recent papers on fallback accretion around neutron stars use a
value of $M_{\rm fallback} \le 0.1 M_{\odot}$ \citep{chat00}.  In either
case, accretion could be significant and might power a central source. In the following sections, we will examine possible accretion scenarios for the central source in SN 1987A.

There are two widely-recognized accretion regimes: a viscous,
differentially rotating flow in which matter loses energy through
viscous friction and spirals in toward the central object, and a
spherically symmetric accretion flow.
For a viscous differentially rotating flow, there are four known
self-consistent solutions to the equations of hydrodynamics,
summarized by \citet{nar98}.  Of these, only two need concern us.  The
first is the famous solution of \citet{sha73}, in which accretion
occurs through a geometrically thin, optically thick disk that
radiates everywhere as a blackbody, with a temperature gradient
throughout the disk.  There are two solutions in the form of
advection-dominated accretion flows (ADAFs), one for optically thick
gas inflow, and one for optically thin, low-density inflow. At the
accretion rates predicted by fallback models (see Figure
\ref{disks_uvcolor}), the
accretion rate is too high to be optically thin.  We
will therefore focus on the optically thick advection-dominated
accretion regime, in the form of a ``slim disk'' model.  The fourth
solution, first proposed by \citet{sha76}, is self-consistent, but
thermally unstable and is therefore a theoretical solution only, not
expected to exist in real systems (as discussed in Narayan,
Mahadevan, \& Quataert 1998).

\subsection{Spherical Accretion} \label{bondi}

In the absence of significant angular momentum, fallback from the
supernova explosion would form a spherical cloud of debris moving radially
toward the central object.  Some models of fallback for SN 1987A assume
spherically symmetric accretion \citep{che89,hou91,bro94}.  \citet{che89} also suggests that there are phases in the accretion process where pressure effects can be neglected, resulting in a ballistic flow.

Throughout the paper, we will parameterize the accretion rate as
\begin{equation} \label{dimmdot}
\dot{m} \equiv \left( \frac{\dot{M}}{\dot{M}_{\rm Edd}} \right) ,
\end{equation}
where $\dot{M}_{\rm{Edd}}$ is the Eddington accretion rate given by
\begin{equation} \label{MEdd}
\dot{M}_{\rm Edd} \equiv \frac{L_{\rm{Edd}}}{\eta_{\rm eff}c^2} ,
\end{equation}
$L_{\rm Edd}$ is the Eddington luminosity, and $\eta_{\rm eff}$ is the efficiency with which gravitational energy is released.
For an object with mass $M$ that is accreting hydrogen gas, 
\begin{equation}
L_{\rm Edd} = 1.30 \times 10^{38} \left( \frac{M}{M_{\odot}} \right)
	\quad\mbox{ergs s}^{-1} .
\end{equation}
The dimensionless luminosity is
\begin{equation} \label{ldef}
l = \left( \frac{L}{L_{\rm Edd}} \right) . 
\end{equation}
Using the value of the efficiency $\eta_{\rm eff} = 0.1$ as given in \citet{fra92} and equation (\ref{MEdd}), the Eddington mass accretion rate is
\begin{equation}
\dot{M}_{\rm Edd} = 2.2 \times 10^{-8} \left( \frac{M}{M_{\odot}} \right)
	M_{\odot} \mbox{ yr}^{-1} . 
\end{equation}

\citet{hou91} use spherical accretion to model neutron-star accretion in SNRs, and then apply their results to SN 1987A.  \citet{bro94} follow the derivation for SN 1987A of \citet{che89}, with the modification of including the neutrino-cooling function derived by \citet{hou91}.  They address scenarios for both neutron-star and black-hole accretion.  Brown {\&} Weingartner estimate that fallback material on the order of 10$^{-4}$--10$^{-3} M_{\odot}$ will remain in the vicinity of the central compact object, after the first 2--3 years of steady accretion.  This material will accrete onto the central object within the next $\sim 10$--10$^3$ years \citep{hou91}.  

If the central object is a neutron star, the accreting matter will
eventually radiate some significant fraction of its rest energy.  
Assuming 10\% radiative efficiency, 10$^{-4} M_{\odot}$ accreting over $\sim$ 
4000 years will result in
radiation on the order of the Eddington luminosity, $L_{\rm Edd} = 1.82
\times 10^{38}$ ergs s$^{-1}$, four orders of magnitude higher than
the observational upper limits.
Dust or no dust, the remnant of SN 1987A
does not harbor a spherically accreting neutron star.

For a black hole, there are mechanisms by which material can accrete
with efficiencies $\ll 0.1$, so accretion at the rate discussed above
does not necessarily lead to Eddington luminosities.  After the
initial 2--3 years of neutrino-dominated steady accretion, the only
energy able to escape the black-hole system is that produced by the
compression work done on the infalling material before it crosses the
photon trapping radius of the black hole. \citet{bro94} show the
trapping radius to be $r_{\rm tr} = 0.6 R \dot{m}$.  Inside this radius,
the time it takes a photon to diffuse outward is larger than the
timescale for advection toward the central object, therefore most of
the radiation would be advected inward and could not escape.  Photon
advection results in luminosities on the order of $L \approx
10^{34}$--$10^{35}$ ergs s$^{-1}$ for accretion rates as high as
$\dot{m} = 10^4$.  The optical limit for SN 1987A at the time of their
paper was $L_{\rm opt} \le 6 \times 10^{36}$ ergs s$^{-1}$, too high to constrain potential black-hole radiation. 

More recently, \citet{zam98} have presented both an analytic analysis and a simulation using a radiation hydrodynamic Lagrangian code to determine the luminosity produced by a black hole undergoing spherical accretion.  Their analytic result is derived by combining the work of \citet{col96}, who solve the equations of hydrodynamics for a fluid accreting onto a stellar remnant from a reverse shock wave, with the \citet{blo86} model of hypercritical spherical accretion.  Applying parameters for SN 1987A given by \citet{che89}, they find
\begin{equation} \label{bhlum}
l = \frac{8\times10^{-3}}{\left(M_{\rm BH}/M_{\odot}\right)^{1/3}
	\left[t\mbox{(yr)}\right]^{25/18}} , 
\end{equation} 
in units of the Eddington luminosity.  When the accretion is optically thick, as modeled here, the numerical
simulations run by Zampieri et al. show both the effective temperature
$T_{\rm eff}$ and the temperature of the gas in the photosphere
$T_{\rm ph}$ to
have values in the range of $\sim 8,000$--10,000 K, producing a roughly
blackbody spectrum.  These temperatures would produce blackbodies with
peaks in the visible to near-UV range, so most of the
radiation should fall within the range we have observed.
The numerical models of {\citet{zam98} are in agreement with the 
above analytical result to within $\sim 30$\%.  We will take this value as an estimate of the 
uncertainty in the model.  

At the time corresponding to our data, $t = 16.75$ years, we find that $l = 1.40 \times 10^{-4}$, 
or $L_{\rm acc} = 2.72 \times 10^{34}$ ergs s$^{-1}$.  Our upper
limit, accounting for 35\% dust absorption, is $L_{\rm opt} \le 8
\times 10^{33}$ ergs s$^{-1}$.  This is about 1/3 the luminosity
predicted by the \citep{zam98} calculation.  

Their calculation is based on a black hole with mass
$M=1.5M_{\odot}$.  It is not clear that black holes form with such low
masses.  Estimates for the lowest possible mass of a black hole range between $1.5M_{\odot}$
and $2.5M_{\odot}$; see \citet{zam98} for a list of these models.  
The corresponding accretion luminosity for a black hole with a mass of 2.5 $M_{\odot}$ is 
$L_{\rm acc} = 3.8 \times 10^{34}$ ergs s$^{-1}$.  \citet{zam98} use the
lowest hypothesized mass limit on a black hole.  Also, \citet{blo86} gives his hypercritical accretion model as a minimum value for the accretion-produced luminosity, based upon the assumption that the only mechanism for accretion-produced luminosity
is the compression of the material outside of the trapping radius.  The calculations of \citet{zam98} produce conservatively low values of accretion radiation, both in the choice of black-hole mass ($M=1.5M_{\odot}$) and in
the assumption that compression of the material outside the trapping radius is the only mechanism for accretion radiation; hence, their result should be a lower limit on the accretion luminosity.  

The lower limit predicted by the Zampieri et al. (1998) model for a 1.5
$M_{\odot}$ black hole undergoing accretion is more than 3 times
larger than our upper limit on a point source in SN 1987A.  In
the case of large amounts of dust absorption ($\ge 80$\%), a
spherically accreting black hole could be concealed within the remnant
of SN 1987A, but in the case of more moderate dust absorption, this
picture conflicts with the observations.

\subsection{Spherical Accretion with Large Iron Opacity} \label{iron}

\citet{fry99} have done fallback accretion modelling in which iron
opacity plays an important role in suppressing accretion.  As the
temperature of the remnant decreases, neutrino cooling falls off and
the radiation is dominated by photon transport.  When the outer layers
of the atmosphere reach the temperature $T = 0.2$ keV, the iron
opacity in the model atmosphere increases by a factor of 300 and dominates over electron scattering.  The iron opacity they
model can be as much as 3--4 orders of magnitude greater than the
opacity caused by Compton scattering, which leads to suppression
factors of $L_{\rm Edd, Fe} / L_{\rm{Edd}, e^-} =
10^{-2.5}$--$10^{-3.9}$.  For singly-ionized, iron-group-rich material, they
use an Eddington limit of
\begin{eqnarray}
L_{\rm{Edd}, e^-} & = & 1.25 \times 10^{38} \left( \frac{M}{M_{\odot}} \right)
\frac{\bar{A}}{f}\mbox{ ergs s}^{-1} \nonumber\\
 & \approx & 7 \times 10^{39} \left( \frac{M}{M_{\odot}}\right) \mbox{ ergs s}^{-1},
\end{eqnarray}
where $\bar{A}$ is the mean atomic weight of the material, and $f$ is
the mean number of free electrons per ion.

With the Eddington limit
suppressed by iron opacity, the opaque outer layer of the atmosphere
prevents higher rates of radiation.  The gravitational energy in the
photons inside this outer layer cannot escape efficiently, and most of
the energy goes into ejecting mass from the system, creating a
photon-driven wind that pushes the outer portions of the atmosphere
away from the stellar remnant while the inner atmosphere is accreted
onto the star at the iron opacity accretion rate.  \citet{fry99}
compute the iron opacity Eddington luminosity at late times for
several values of atmosphere density, by applying these values of the
suppression factor to the Eddington limit for singly-ionized
iron-rich material.  By 5000 days (13.7 years) past the SN
explosion, they find the iron opacity Eddington luminosity levelling
off.  Extrapolating the
curves shown in their Figure 8 to the time of our observations at day
6110, we find $L_{\rm{Edd},Fe} = 2.0 \times 10^{36}$--$1.5 \times
10^{37}$ ergs s$^{-1}$ for densities in the range $10^{-8} \le \rho
\le 10^{-6}$ g cm$^{-3}$.  
These values are 2--3 orders of magnitude higher than our upper limit,
and are therefore inconsistent with the observations.  If this model
is relevant for SN 1987A, the photon-driven wind must have removed the
entire envelope of accreting material on a timescale of less than 
12 years, the time of our spectral observations.

\subsection{Thin Disk Models} \label{disk}

Spherically symmetric accretion may not be the best approximation of
accretion onto the core of SN 1987A.  The ring structures, polarization, shape of the debris and other aspects of the explosion all indicate a special axis that could be an axis of rotation \citep{wan02}. 
If there is a significant amount of angular momentum in the infalling
material, the matter will settle into a geometrically thin disk, with each
part of the disk orbiting the central source in a Keplerian orbit.  

In systems where the timescale of external change (i.e., the mass infall) is greater than the timescale of viscosity spreading within the disk, the accreting matter settles into a steady-state disk.  Steady-state disks are modeled to have a viscous dissipation flux $D(R)$ that does not depend on the local viscosity of the disk, but only on the mass of the central object $M$, and the accretion rate $\dot{M}$, as given in \citet{sha73}: 
\begin{equation}
D(R) = \frac{3GM\dot{M}}{8 \pi R^3} \left[ 1- \left( \frac{R_{\ast}}{R} \right)
	^{1/2} \right] ,
\end{equation}
where $R_{\ast}$ is the radius of the neutron star.
If we make the assumption that the disk is optically thick in the $z$-direction, the disk will radiate as a series of concentric blackbody rings, each with an effective temperature given by the standard blackbody relation, $\sigma T^4 = D(R)$, and
\begin{equation} \label{disktemp}
T(R) = \left\{ \frac{3GM\dot{M}}{8 \pi R^3 \sigma}\left[ 1- \left( 
	\frac{R_{\ast}}{R} \right)^{1/2} \right] \right\}^{1/4} . 
\end{equation}
Therefore the flux per unit wavelength radiated from the disk is given
by integrating the blackbody surface flux over the surface area of the disk,
\begin{equation}
F_{\lambda} = \int\limits^{R_{\rm max}}_{R_{\rm min}} \frac{2\pi hc^2}{\lambda^5} \left[
	\frac{1}{\rm{exp}[hc/{\lambda}kT(R)]-1} \right] 4 \pi R dR .
\end{equation}

We computed the flux from a thin disk by approximating a thin accretion
disk as a set of concentric blackbody rings of constant temperature, using
the temperature determined by equation (\ref{disktemp}).  We integrated across the disk
surface to compute the flux at each wavelength, and then produced a model
spectrum for the disk.  We took the
inner radius of the disk to be the radius of a neutron star (that is,
$R_{\rm min} = 1 \times 10^6$ cm), the mass of the central object to be $M =
1.4 M_{\odot}$, and the distance to the object to be 51.4 kpc.  
The model spectra were reddened as in \S \ref{image} and converted to
a total count level using the {\it synphot} task {\it calcspec}.  The
total counts were then compared to the upper limit from \S \ref{image}
for each filter to determine which disks could hide in the SNR. 

Because the high
temperatures in the disk near the compact object produce X-ray and far-UV radiation,
our model for the near-UV and optical spectrum is insensitive to the exact
value of $R_{\rm min}$ (values of $R_{\rm min} = 10$ km and $R_{\rm min} = 300$ km
gave identical output to within 1\%).  As seen in equation (\ref{disktemp}) the
temperature scales as $(M\dot{M})^{1/4}$, allowing the mass difference
between a 1.4 $M_{\odot}$ neutron star and a 1.5 $M_{\odot}$ black hole to
be absorbed into the computed value of $\dot{M}$.  That is, the value of
$\dot{M}$ for a 1.5 $M_{\odot}$ black hole accretion system at a given
value of $R_{\rm max}$ will correspond to $1.4/1.5=0.93$ times the
value of $\dot{M}$ for a 1.4 $M_{\odot}$ neutron star system with the same
$R_{\rm max}$.  The model was run with varying values of the outer radius
$R_{\rm max}$ and the dimensionless accretion rate $\dot{m}$ (see equation
\ref{dimmdot}).  

Figure \ref{disks_uvcolor}a shows the values of $R_{\rm max}$ and $\dot{m}$
that are consistent with our upper limits in each band.  The UV filter F330W (solid curve) gave
the most stringent limit on $R_{\rm max}$ and $\dot{m}$.  The
shaded region represents disk parameters that are excluded by our
UV upper limit.  From the figure, we can see that an accretion disk in
the SNR is limited to small disk sizes and low accretion rates.  The
restrictions in $R_{\rm max}$-$\dot{m}$ space given by other filter bands
are also shown: the dotted curve is the limit from the B-band filter
(F435W), the short-dashed curve corresponds to the V-band filter (F555W), the
dot-dashed curve to the R-band filter (F625W), and the long-dashed curve
to the I-band filter (F814W).  We ran the simulation out to $R_{\rm max} =
1 \times 10^{12}$ cm.  Beyond this point, the disk is cool and
contributes only in the IR, therefore our optical constraint on
$\dot{m}$ will remain constant beyond $R_{\rm max} = 1 \times 10^{12}$ cm,
as can be seen in Figure \ref{disks_uvcolor}.

Any accretion disk in SN 1987A must be below the luminosity limit in
all bands to escape detection.  The UV-band limit is the most
stringent, so we have used it to delineate the part of
$R_{\rm max}$-$\dot{m}$ space that is consistent with our upper limit.
Uncertainties in the correction for interstellar reddening raise the
UV-band limiting curve by at most a factor of two, and therefore
do not qualitatively change our results for thin disk models.

It is worth investigating how various levels of dust absorption within
the remnant change the region of $R_{\rm max}$-$\dot{m}$ space that is
consistent with our limit.  Figure \ref{disks_uvcolor}b shows the
excluded region for various levels of dust absorption.  These limits
are derived from the UV-band limit, since it gives the most stringent
constraint.  The excluded region in the case of no dust absorption is
the solid curve from Figure \ref{disks_uvcolor}a.  Three other cases are shown:
a system in which 35\% of the flux is absorbed by dust, 70\% of the
flux is absorbed by dust, and 97\% of the flux is absorbed by dust.
Recall that 35\% is the predicted amount of dust absorption based on
the argument in \S \ref{dust}, so this line corresponds to our upper
limit of $L_{\rm opt} \le 8 \times 10^{33}$ ergs s$^{-1}$, and is the
value we will use in our analysis.  The case of 70\% dust absorption would
apply if our estimate of the internal dust absorption is off by a
factor of two, and 97\% absorption corresponds to the
worst-case-scenario of a single opaque dust clump along the line of
sight to the remnant.

For this limit to be useful, we must have some understanding of what
``typical'' values for $R_{\rm max}$ and $\dot{m}$ are.  Standard
values of $R_{\rm max}$ for binary systems are given as (2--8)$\times 10^{10}$ cm in 
\citet{bat81}, and of order $10^{10}$ cm in \citet{fra92}.  (These do not necessarily apply 
to accretion disks in SNRs, but will we use this range as an estimate of the typical size 
of an accretion disk around a stellar remnant.)  For a disk of radius
$R_{\rm max} = 1 \times 10^{10}$ cm, our observations limit the accretion
rate to $\dot{m} \le 0.3$, in the case of 35\% dust absorption.  In \S\ref{fall}, we will compare
out limits on disk size and accretion rate to the values for $R_{\rm max}$
and $\dot{m}$ predicted by fallback models.

\subsection{The Slim Disk Model} \label{slimdisk}

Thin disk models operate under the assumption that the vertical
structure of the disk can be treated separately from the radial
structure, but this assumption breaks down at high (super-Eddington)
accretion rates, ie. where $\dot{m} \ge 1$.  Beyond a critical value
of the accretion rate, the disk must have non-negligible thickness to
remain in hydrostatic equilibrium.  An early description of these
slim disks is given in \citet{jar80}.  They show that disks with
super-critical accretion rates have luminosities near or above
$L_{\rm{Edd}}$ which radiate predominantly in X-rays.  For a central
source with $M = 1.4 M_{\odot}$, the critical accretion rate of
\citet{jar80} is a few times $10^{-8} M_{\odot}$ yr$^{-1}$, comparable to the Eddington accretion
rate $\dot{M_{\rm{Edd}}}$ we defined in \S\ref{bondi}.  In the next
section, we will see that models of fallback in SNRs predict accretion
rates after $\sim 16.75$ years of order the Eddington accretion rate,
with $\dot{m} \sim 1$, as shown by the red and blue grids in Figure
\ref{disks_uvcolor}.  At accretion rates this high, the assumption of
a thin disk begins to break down.  However, we would expect to see
accretion luminosities of order $L_{\rm{Edd}} \sim 2 \times 10^{38}$
ergs s$^{-1}$ for a slim disk with an accretion rate this high.  These luminosities are 4 orders of magnitude higher than
both the optical upper limits for SN 1987A derived in this paper, and
the X-ray upper limit of \citet{par04}, therefore
it appears unlikely that super-Eddington, slim disk accretion is an
appropriate model for SN 1987A. 

\section{Fallback Models for SNRs} \label{fall}

We have used our upper limit on the continuum flux from SN 1987A to
constrain the rate of accretion onto a compact object under standard
spherical, thin disk, and slim disk accretion models.  It is instructive to compare these limiting rates to models for the fallback onto a SNR.  In studying anomalous X-ray pulsars (AXPs), \citet{chat00} have developed a model for neutron-star accretion from a disk of fallback debris in the aftermath of a supernova explosion.  Their model examines accretion from this ``fossil disk'' for various values of the magnetic field $B$ and initial rotation period $P_0$ of the neutron star, and the mass of the disk $M_d$.  They find that for some values of these parameters (specifically $B \le 3.9 \times 10^{12}$ Gauss), no efficient accretion mechanism evolves and the neutron star becomes a radio pulsar on a timescale of $\sim 100$ years.  For higher values of $B$, the accreting neutron star becomes an AXP.  From a set of initial conditions, they evolve a time-dependent expression for the disk accretion rate $\dot{M}(t)$ as a function of time $t$ from the explosion.  The \citet{chat00} model has been used by Perna, Hernquist, \& Narayan (2000; hereafter PHN) and later by Menou, Perna, \& Hernquist (2001; hereafter MPH) to develop more detailed models of the fallback onto a compact object in a SNR.  The two models differ in their choice of timescale for fallback.  In this section, we compare the predicted fallback accretion in both of these models with the limits on $\dot{m}$ for various accretion scenarios discussed in \S \ref{acc}. 

\subsection{Fallback Models with a Steady-State Disk} \label{fallmod}

Both the PHN and MPH fallback models are based on
the \citet{pri74} solution to a steady-state accretion disk.  They use
the self-similar solution for the Pringle disk found by \citet{can90}:
\begin{equation} \label{selfmass}
M_d(t) = M_d(t_0) \left( \frac{t}{t_0} \right)^{-3/16} ,
\end{equation}
\begin{equation} \label{selfrad}
R_d(t) = R_d(t_0) \left( \frac{t}{t_0} \right)^{3/8} ,
\end{equation}
where $t_0$ is the timescale for disk formation, $M_d$ is the total
disk mass, and $R_d$ is the outer radius of the disk.  From equation
(\ref{selfmass}) it follows that 
\begin{equation} \label{selfmdot}
\dot{M}_d(t) = - \frac{3}{16} M_d(t_0) \left( \frac{t}{t_0} \right)^{-3/16}
	\left( \frac{1}{t} \right) .
\end{equation}

We can use our limits for the $\dot{M}$ and $R$ of a thin disk from \S \ref{disk} in conjunction with the initial conditions of the disk, $M_d(t_0)$ and $R_d(t_0)$, to determine whether or not the proposed fallback models are compatible with our observed upper luminosity limit.  In doing
so, it is useful to have a general idea of likely values for
these conditions.  The total fallback material accreted onto the
compact object is predicted to be $M_{\rm fallback} \le 0.1$--$0.15
M_{\odot}$ \citep{che89,chat00}.  Of this, a fraction will settle into
the initial accretion disk.  PHN use a value of $M_d =
0.005 M_{\odot}$.  We will take a broad range of $10^{-5} M_{\odot}
\le M_d(t_0) \le 0.1 M_{\odot}$ as a reasonable estimate for the initial mass of the
accretion disk.  For the initial radius, MPH argue that the fallback material
originates outside of the core of the progenitor star with $R \le
10^9$ cm.  Using predictions for the angular momentum of the fallback
material (from simulations by Heger, Langer, \& Woosley 2000) and conserving angular
momentum, they calculate Keplerian orbits in the range $10^6$ cm $ \le
R_d(t_0) \le 10^8$ cm.

To use equations (\ref{selfmass}--\ref{selfmdot}), we will need an expression
for $t_0$.  This is where the PHN and MPH models differ.  PHN assume
that a fallback disk will form on the order of milliseconds.  They use a
value of $t_0$ = 1 ms, following \citet{chat00}, whose numerical
calculations show the models to be relatively insensitive to $t_0$
over larger timescales.  MPH argue that disk formation ought to occur
on the local viscous timescale, on the order of $10^3$ s. Although
$M_d$ and $R_d$ are only weakly dependent on $t_0$ ($M_d \propto
t_0^{3/16}$ and $R_d \propto t_0^{-3/8}$), the choice of timescale
turns out to be significant over the period of 16.75 years that have elapsed between the SN 1987A explosion and our observations. 

\subsection{The PHN Model} \label{phn}

Using equations \ref{selfrad} and \ref{selfmdot}, with a
characteristic timescale of $t_0$ = 1 ms, we can calculate a set of
disk sizes and accretion rates predicted by the PHN model for the
time of our observations, 16.75 years after outburst.  The range of
predicted $R_{\rm max}$ and $\dot{m}$ are shown in Figure
\ref{disks_uvcolor}a as the blue grid.  The vertical lines correspond to
initial disk radii (from left to right) of $10^6$ cm, $10^7$ cm, and
$10^8$ cm.  The horizontal lines correspond to initial disk masses
(from top to bottom) of $10^{-1} M_{\odot}$, $10^{-2} M_{\odot}$,
$10^{-3} M_{\odot}$, $10^{-4} M_{\odot}$, and $10^{-5} M_{\odot}$.  We
can see from Figure \ref{disks_uvcolor}a that the PHN model is not consistent
with our upper limits.  If there is substantial dust absorption in the
SNR ($\ge 70$\%) then the smallest and least massive initial
disks predicted by the PHN model may be consistent with our limits.
In general, the PHN model predicts higher accretion rates than are
compatible with our observed upper limit.

According to the \citet{chat00} model, neutron-star systems enter a phase in which the
rapid rotation of the star's magnetic field acts as a ``propeller''
that throws matter out from the surface of the star.  In this
scenario, matter cannot accrete onto the star.  The propeller effect
was first described by \citet{ill75}.  They define the corotation
radius $R_{\rm co}$ as the radius at which the accretion disk rotates with
the same angular frequency as the neutron star.  The magnetospheric
radius $R_{\rm mag}$
is the radius at which magnetic forces dominate over the gravitational
force of the neutron star. When $R_{\rm co} < R_{\rm mag}$, the magnetosphere is rotating
fast enough that the centrifugal force overcomes the gravitational
pull of the neutron star and the accreting material is pushed outward.  Although
it is unclear whether the material is actually expelled from the
system, or remains outside of the magnetosphere, it is unable to
accrete onto the neutron star in either case \citep{men99}.  The propeller
effect can operate in a thin-disk accretion regime, where the
propeller effectively prevents any accretion, or in a spherical ADAF
regime, where some matter is able to accrete along the poles of the
neutron star's rotational axis.  Can the existence of a propeller around a neutron star in SN
1987A account for the low-level of accretion observed, as
compared to the predicted level in PHN? 

The propeller is only effective inside the magnetospheric radius of the
neutron star where it prevents accretion onto the surface of the neutron
star.  Outside of the magnetospheric radius, the thin disk
surrounding the propeller will still be able to radiate away some of the
gravitational energy stored in the accretion flow.  The magnetospheric
radius for a neutron star is given by \citet{fra92} as 
\begin{equation}
\label{rmag} R_{\rm mag} = 6.6 \times 10^7 B^{4/7}_{12} \dot{m}^{-2/7}\mbox{cm}, 
\end{equation} 
where $B_{12}$ is the magnetic field of the
neutron star expressed in units of $10^{12}$ Gauss.  Only large values of
$R_{\rm mag}$ will affect the optical flux from the disk (recall from \S
\ref{disk} that the inner portion of the disk contributes little flux in
the optical band), therefore we maximize the propeller effect by choosing
a large value for the pulsar magnetic field, $B=10^{12}$ Gauss.  When we
repeated our thin accretion disk simulations for a $1.4 M_{\odot}$ neutron
star with a magnetic field of $10^{12}$ Gauss and an inner disk radius
given by equation (\ref{rmag}), we found that the resulting disk
luminosities were comparable to those of disks without a propeller.  The
presence of a propeller does not substantially change the optical
luminosity of a given size disk, so the accretion rate and
outer radius limits we found for non-propeller systems apply to propeller
systems as well. The PHN propeller cannot explain the low level of
optical emission observed from SN 1987A.

\subsection{The MPH Model} \label{mph}

In the MPH model, the authors assume that the fallback material will settle into a
disk on a timescale determined by the local viscous timescale.
They write
\begin{eqnarray} \label{timescale}
t_0 & \equiv & \frac{R^2 \Omega _K}{\alpha c^2_S} \nonumber\\
 & \approx  & 6.6 \times
	10^{-5} \left( \frac{T_i}{10^6 \mbox{K}} \right)^{-1} 
	\left( \frac{R_d(t_0)}{10^8 \mbox{cm}} \right)^{1/2} \mbox{yr},
\end{eqnarray}
where $\Omega_K$ is the local Keplerian angular speed, $c_S$ is the
speed of sound, and $T_i$ is the initial temperature of the newly formed disk.  This equation is based on an
assumed viscosity paramter of $\alpha = 0.1$ and a compact object with
$M = 1 M_{\odot}$.  A range of temperatures for young accretion disks is
given by \citet{ham98} as $10^4$ K $\le T_i \le 10^6$ K.

We can use equations (\ref{selfrad}) and (\ref{selfmdot}) in
conjunction with reasonable initial values for $R_d(t_0)$, $M_d(t_0)$,
and $T_i$ to compare the predictions of the MPH model with our upper
limits for accretion-disk radiation as calculated in \S \ref{disk}.
As in \S\ref{phn}, we investigated the following range of initial
values for the radius, and mass of the accretion disk:

\begin{equation}
10^6 \mbox{ cm} \le R_d(t_0) \le 10^8 \mbox{ cm}
\end{equation}
\begin{equation}
10^{-5} M_{\odot} \le M_d(t_0) \le 10^{-1} M_{\odot}.
\end{equation}

We used an initial temperature range of

\begin{equation}
10^4 \mbox{ K} \le T_i \le 10^6 \mbox{ K}.
\end{equation}

The ranges of predicted disk radii and accretion rates at $t=16.75$
years after outburst are shown as the red grids in Figure
\ref{disks_uvcolor}.  The solid grid is for disks with temperature $T =
10^6$K; the darker dashed grid is for disks with $T=10^4$K.  As before, the
vertical lines are for initial disk radii (from left to right) of
$10^6$ cm, $10^7$ cm, and $10^8$ cm, while the nearly-horizontal lines
are for initial disk masses (from top to bottom) of $10^{-1}
M_{\odot}$, $10^{-2} M_{\odot}$, $10^{-3} M_{\odot}$, $10^{-4}
M_{\odot}$, and $10^{-5} M_{\odot}$.  

Our upper limit provides some restriction on the radius and accretion
rate of a hot ($T=10^6$K) disk, limiting a disk in the central debris
of SN 1987A to a small, disk ($R_{\rm max} \le 1 \times 10^{10}$ cm) with a
low accretion rate.  For disks with initial masses of $M \ge 0.005
M_{\odot}$, the disks must be even smaller ($R_{\rm max} \le 5 \times
10^9$).  For cooler
initial temperatures, disks of the same initial radius do not grow as
large as comparable disks at hotter temperatures.  Only the largest and
most massive of the cooler disks are restricted by our observed upper
limits.

Notice that the MPH model predicts disk sizes that are smaller than
the standard values of (2--8)$\times 10^{10}$ cm for accretion disks
in binary systems.  Disks with $R_{\rm max}$ larger than $\sim 1\times
10^{10}$ cm are incompatible with our limits on $\dot{m}$ unless they
have less accretion than predicted by these fallback models.  The
smaller, cooler disks of the MPH model fit within our observed upper
limits.

It is worth noting that the predictions of the angular momentum given in
\S\ref{fallmod} for fallback systems may be too large.  Recent simulations by
\citet{heg04} include the effects of magnetic fields in the end stages
of nuclear burning in massive stars.  They find that a significant
amount of angular momentum is lost through stellar winds, and that
angular momentum is transferred from the core of the star outward
through magnetic braking.  Consequently, the material around the iron
core prior to collapse has values of the specific angular momentum $j$
that are 30--50 times lower than those found in the absence of
magnetic fields.  For Keplerian disk orbits, $j \approx \sqrt{GMr}$,
where $M$ is the enclosed mass of the orbit, and $r$ is the radius of
the orbit.  A decrease in $j$ by a factor of 30--50 corresponds to a
decrease in the outer radius of the disk by about three orders of
magnitude, giving initial disk radii of $10^3$ cm $ \le R_d(t_0) \le
10^5$ cm.  These are non-sensical values for the radius of a disk
around a neutron star, since they are smaller than the outer radius of
the star: $R_{NS} \sim 10$ km.  For a neutron star, initial disk radii of
$R_d(t_0) \sim 10^6$--$10^7$ cm are the smallest possible disk sizes.  The low
values of angular momentum predicted by \citet{heg04} suggest that
the smallest disks are the most likely, or that an accretion disk may
not form at all.

\section{Conclusions} \label{conc}

Taking advantage of the high resolution and sensitivity of {\it
ACS/HRC} on {\it HST}, we have measured the lowest upper limit on the
optical and near-UV luminosity to date for a point source at the
center of the SN 1987A remnant.  We find that the total optical
flux is limited to $F_{\rm opt} \le 1.6 \times 10^{-14}$ ergs s$^{-1}$
cm$^{-2}$.  The luminosity limit can be written

\begin{eqnarray}
\lefteqn{L_{\rm opt} \le 5 \times 10^{33} \mbox{ergs s}^{-1}}\nonumber\\
& & \times \left(
\frac{D}{51.4 \mbox{ kpc}} \right)^2 \times \mbox{exp} \left[
  \frac{A_V}{0.595} \right] \times \left( \frac{1}{1-\alpha_0}
\right) ^{ \left( t_0/t \right)^{2}}
\end{eqnarray}
where $D$ is the distance to the supernova remnant, $A_V$ is the
actual value of the reddening in the direction of SN 1987A, $t$ is the
time since outburst, $t_0$ is the epoch of
dust formation, and $\alpha_0$ is the fraction of the luminosity that is
absorbed within the remnant at time $t_0$.  Writing the
luminosity limit in this form takes into account the effects of
distance, reddening, and internal dust absorption, and of the
uncertainties in the upper limit due to these effects.  We have used $t_0$ = 2172 days and $\alpha_0 = 0.97$, resulting in an
upper limit of $L_{\rm{opt}} \le 8 \times 10^{33}$ ergs s$^{-1}$ for a
point source in the remnant of SN 1987A at $t \approx 6110$ days.

We have also measured upper limits on a continuum source in
the SNR by fitting a continuum level to the flux between emission
lines in the spectrum of SN 1987A from 1999 September.  The spectral
data give limits on the
luminosity of $L_{\rm UV} \le 1.7 \times 10^{34}$ ergs s$^{-1}$ and
$L_{\rm opt} \le 1.9 \times 10^{34}$ ergs s$^{-1}$.  These limits are not
as stringent as those derived from the more recent imaging data,
suggesting that the remnant has dimmed in the intervening time.  All of these limits are likely to be lower than the 
actual luminosity limit on a compact remnant, due to dust absorption within the SNR itself.  In the 
worst-case scenario for dust absorption, as much as 97\% of the light emitted by a central source may 
be absorbed by a thick clump of black dust along the line of sight to
the SNR.

Based upon our observed upper limit, we find that a possible surviving binary 
companion to SN 1987A is limited to F6 and later stars.  The limits on
a point source in the SNR require a potential pulsar to have either a
very weak magnetic dipole moment, or a very strong magnetic field that
puts the magnetized neutron stars in the realm of magnetars.  The
pulsar must be fainter in X-rays than the other young pulsars with ages
less than 10,000 years known to be associated with SNRs.

We have examined the available models for accretion scenarios.
We find that spherical accretion onto a neutron star is ruled out by a factor of
$10^4$.  A spherically accreting black hole is also incompatible with
our limit, unless the dust absorption exceeds 80\%.  We have also
examined accretion scenarios in which the fallback material
has significant angular momentum and forms a thin accretion disk or an
optically thick slim disk.  In the case of a thin disk, we have shown that the accretion
disk must have a small radius to correspond with predicted values of
the accretion rate $\dot{m}$, for either a neutron star or a black hole.  
The presence of a magnetized propeller does not
substantially affect this result.  The \citet{men01} fallback model is
consistent with the presence of a small disk accreting at a rate
within our measured upper limit.  A slim disk is incompatible with our
upper limit.  Other possibilities include the scenario in which photon winds have driven all
the remaining fallback material out of the system, truncating
accretion, as proposed by \citet{fry99}, or that all of the fallback
material has already accreted onto the compact remnant on a timescale
of less than 14 years.

Future spectroscopic observations in the mid- to far-IR with the {\it Spitzer Space Telescope} may 
provide additional data on the dust remaining in 
the SNR.  A better determination of the mechanisms which power the
10$\mu$m radiation from the central ejecta might make it possible to
determine the extent to which optical luminosity from a central source
is being absorbed by dust in the remnant.  Also, as the remnant 
expands, the dust should thin, allowing us to peer more deeply into the remnant. Over time, as the 
radiation from debris in the remnant grows dimmer and dust absorption declines, we will be able 
to place increasingly strong limits on a point source in SN 1987A.  Eventually a point source 
may even be detected!

\acknowledgments

The authors would like to thank Thomas Matheson, Ramesh Narayan, and Saurabh
Jha for helpful discussion and suggetions.  Support for HST proposal
number GO 09428 was provided by NASA through a grant from the Space
Telescope Science Institute, which is operated by the Association of
Universities for Research in Astronomy, Inc., under NASA contract
NAS5-26555.  RAC acknowledges support from NASA grant NAG5-13272.  AVF
is grateful for a Miller Research Professorship at U. C. Berkeley,
during which part of this work was completed.

\clearpage

\begin{deluxetable}{cccc}
\tabletypesize{\scriptsize}
\tablecaption{STIS Data. \label{stis}}
\tablewidth{0pt}
\tablehead{
\colhead{Grating} & 
\colhead{Slit}   & 
\colhead{Exposure Time}   &
\colhead{Wavelength Range}\\
\colhead{} &
\colhead{} &
\colhead{(s)} &
\colhead{({\AA})}
}
\startdata
G140L &$0.5''$ &10400  &1140--1730\\
G230L &$0.5''$ &10400  &1568--3184\\
G430L &$0.2''$ &7800  &2900--5700\\
G750L &$0.2''$ &7800  &5236--10266\\
\enddata
\end{deluxetable}

\clearpage

\begin{deluxetable}{ccccc}
\tabletypesize{\scriptsize}
\tablecaption{ACS Imaging Data. \label{acsdata}}
\tablewidth{0pt}
\tablehead{
\colhead{ACS/HRC} &
\colhead{Band} &
\colhead{Exposure Time}   &
\colhead{Pivot Wavelength} &
\colhead{RMS Bandwidth} \\
\colhead{Filter} &
\colhead{} &
\colhead{(s)} &
\colhead{({\AA})} &
\colhead{({\AA})}
}

\startdata
F330W &HRC UV &400 &3362.7 &1738.2  \\
F435W &Johnson B &400 &4311.0 &3096.8  \\
F555W &Johnson V &200 &5355.9 &3571.9  \\
F625W &SDSS r &200 &6295.5 &4153.1  \\
F814W &Broad I &200 &8115.3 &7034.5  \\
\enddata

\end{deluxetable}

\clearpage

\begin{deluxetable}{ccccccc}
\tabletypesize{\scriptsize}
\tablecaption{Limits From ACS Images. \label{filterlimits}}
\tablewidth{0pt}
\tablehead{
\colhead{} &
\colhead{Flux Limit of} &
\colhead{Flux Limit of} &
\colhead{Effective $F_{\lambda}$} &
\colhead{Applied Filter} &
\colhead{} &
\colhead{} \\
\colhead{Filter} &
\colhead{4-pixel Aperture} &
\colhead{Infinite Aperture} &
\colhead{(10$^{-18}$ ergs } & 
\colhead{Width} & 
\colhead{$L$} &
\colhead{$L$} \\
\colhead{} &
\colhead{(counts)} &
\colhead{(counts)} &
\colhead{s$^{-1}$ cm$^{-2}$ {\AA}$^{-1}$)} &
\colhead{({\AA})} &
\colhead{(10$^{32}$ ergs s$^{-1}$)} &
\colhead{($L_{\odot}$)}
}

\startdata
F330W &0.51 &0.71 &3.97 &2900--3690 &9.96 &0.26 \\
F435W &2.55 &3.27 &3.59 &3690--4800 &12.7 &0.33 \\
F555W &2.93 &3.80 &1.99 &4800--5700 &5.96 &0.15 \\
F625W &6.62 &8.67 &2.70 &5700--7050 &11.6 &0.30 \\
F814W &4.86 &7.61 &1.32 &7050--9650 &10.9 &0.28 \\
\enddata

\end{deluxetable}

\clearpage

\begin{deluxetable}{llccccc}
\tabletypesize{\scriptsize}
\tablecaption{Comparison with Point Sources in Other SNRs\label{compare}}
\tablewidth{0pt}
\tablehead{
\colhead{} &
\colhead{} &
\colhead{log $L_X$} &
\colhead{log $L_{\rm opt}$} &
\colhead{Age} &
\colhead{Possible in} &
\colhead{} \\
\colhead{SNR} &
\colhead{Source} &
\colhead{(ergs s$^{-1}$)} &
\colhead{(ergs s$^{-1}$)} &
\colhead{(yr)} &
\colhead{SN 1987A?} &
\colhead{Reference}
}

\startdata
SN 1987A  &Point source   &$\le33.74$   &$\le33.9$   &16.75 &   &1,2   \\[3pt]
\tableline \\
\multicolumn{6}{c}{Young Pulsars}   \\[3pt]
\tableline \\
Kes 75   &PSR J1846-0258   &$>34.6$   &... &1700   &N
&3\tablenotemark{*}, 27  \\
Crab   &PSR B0531+21   &36.2   &33.8   &950  &N   &4,
5\tablenotemark{\dag}, 28  \\
N158A   &PSR B0540-69   &36.4   &33.9   &1660 &N
&6, 7\tablenotemark{\dag}, 30   \\
N157B   &PSR J0537-6910   &35.5   &$\le$33.1  &5000  &N
&8\tablenotemark{*}, 25, 31 \\
MSH 15-52   &PSR B1509-58   &35.3   &...   &1800  &N
&9\tablenotemark{*}, 32   \\
Vela   &PSR B0833-45   &31.3   &28.8   &$1.1\times10^4$  &Y
&4, 10\tablenotemark{\dag}, 29   \\
Monogem Ring   &PSR B0656+14   &30.2   &28.2  &$1.1\times10^5$  &Y
&4, 11-13\tablenotemark{\dag}, 29   \\
Geminga   &PSR J0633+1746   &30.2   &27.5   &$3.4\times10^5$ &Y
&4, 11, 13\tablenotemark{\dag},  29   \\[3pt]
\tableline \\
\multicolumn{6}{c}{Nonplerionic X-Ray Point Sources in SNRs}   \\[3pt]
\tableline \\
Cas A   &Point source   &33.8-34.6/33.3\tablenotemark{a}
&$\le$29.1\tablenotemark{\ddag}   &400 &N/Y
&14\tablenotemark{*}, 26, 34   \\
Pup A   &1E 0820-4247   &33.6\tablenotemark{b}
&$\le$30.3\tablenotemark{\ddag}   &3000  &Y
&15\tablenotemark{*}, 27, 33  \\
RCW 103   &1E 1614-5055   &33.9\tablenotemark{b}
&$\le$30.8\tablenotemark{\ddag}   &8000  &N   &16\tablenotemark{*},
27, 35   \\
PKS 1209-52   &1E 1207-5209   &33.1\tablenotemark{b}
&$\le$30.1\tablenotemark{\ddag}   &7000  &Y
&17, 18\tablenotemark{*}, 27, 36   \\[3pt]
\tableline \\
\multicolumn{6}{c}{Anomalous X-ray Pulsars}   \\[3pt]
\tableline \\
Kes 73   &1E 1841-045   &35.5   &...   &$\le2000$  &N
&19\tablenotemark{*}, 37  \\
G29.6+0.1   &AX J1845-0258   &38.6/34.9\tablenotemark{a}   &...
&$\le8000$  &N   &20\tablenotemark{*}, 38   \\
CTB 109   &1E 2259+586   &36.9\tablenotemark{c}   &...  &8800  &N
&21\tablenotemark{*}, 39   \\[3pt]
\tableline \\
\multicolumn{6}{c}{Soft Gamma Repeaters}   \\[3pt]
\tableline \\
G42.8+0.6?   &SGR 1900+14   &34.6\tablenotemark{c}   &...   &$10^4$ &N
&22\tablenotemark{*}, 40   \\
G337.0-0.1?   &SGR 1627-41   &35.8   &...   &5000 &N   &23, 24\tablenotemark{*}  \\[3pt]
\enddata
\tablecomments{X-ray luminosities are in the 1-10 keV band.  All X-ray luminosities log $L_X$ are for power law sources,
  except as noted.}
\tablenotetext{a}{X-ray data can be fit by either a power law or a
  blackbody source, shown log $L_{\rm pl}$/log $L_{\rm bb}$.}
\tablenotetext{b}{X-ray luminosity is for a blackbody source.}
\tablenotetext{c}{X-ray luminosity includes contributions from both
  power law and blackbody models.}
\tablenotetext{*}{Taken from Table 3 in \citet{cha01}.}
\tablenotetext{\dag}{Taken from Table 2 in \citet{zav04b}.}
\tablenotetext{\ddag}{Optical upper limits were computed from {\it R}-band
  limiting magnitudes (Johnson {\it R} in the case of Cas A, SDSS {\it r} in the
  case of Pup A, RCW 103, \& PKS 1209-52).  An optical power law
  spectrum was assumed with a negative spectral index $\alpha =$ -1.07, based on the optical
  spectrum of PSR B0540-69 \citep{ser04}, and integrated over the
 wavelength range 2900--9650\AA.}
\tablerefs{(1) \citealt{par04}. (2) This paper. (3)
  \citealt{got00}. (4) \citealt{zav04b}. (5) \citealt{sol00}. (6)
  \citealt{kaa01}. (7) \citealt{mid87}. (8) \citealt{wan98}. (9)
  \citealt{mar97}. (10) \citealt{shi03}. (11) \citealt{zav04a}. (12)
  \citealt{pav02}. (13) \citealt{kop01}. (14) \citealt{cha01}. (15)
  \citealt{pet96}. (16) \citealt{mer96}. (17) \citealt{tuo80}. (18)
  \citealt{got97}. (19) \citealt{got99}. (20) \citealt{tor98}. (21)
  \citealt{rho97}. (22) \citealt{wood99}. (23) \citealt{cor99}. (24)
  \citealt{hur00}. (25) \citealt{fes02}. (26) \citealt{wanc02}. (27)
  \citealt{mig00}. (28) \citealt{mer02}. (29) \citealt{pav96}. (30)
  \citealt{sew84}. (31) \citealt{mar98}. (32) \citealt{tho92}. (33)
  \citealt{win86}. (34) \citealt{van83}. (35) \citealt{torii98}. (36)
  \citealt{pavlov02}. (37) \citealt{gotthelf97}. (38) \citealt{gae99}.
  (39) \citealt{sas04}. (40) \citealt{vas94}.}

\end{deluxetable}

\clearpage

\begin{figure} 
\epsscale{1}
\plotone{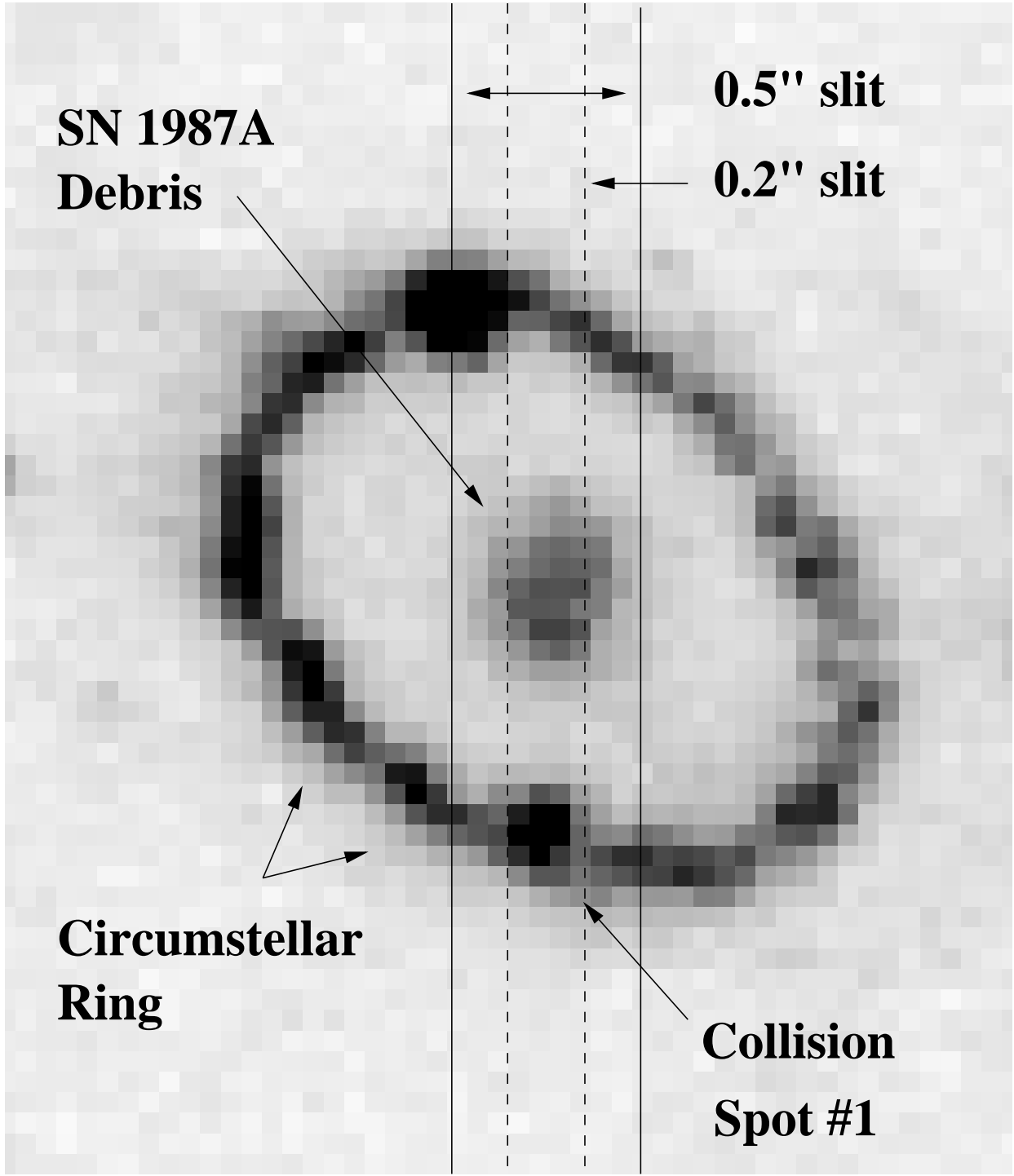}
\caption{{\it STIS} slit positions for 1999 spectral data include the central debris and 
	portions of the circumstellar ring. The $0.5''$ slit was used for the UV spectrum and
	the $0.2''$ slit was used for the optical spectrum (see Table \ref{stis}).}  \label{slitpos}
\end{figure} 

\clearpage  

\begin{figure}
\epsscale{0.85}
\plotone{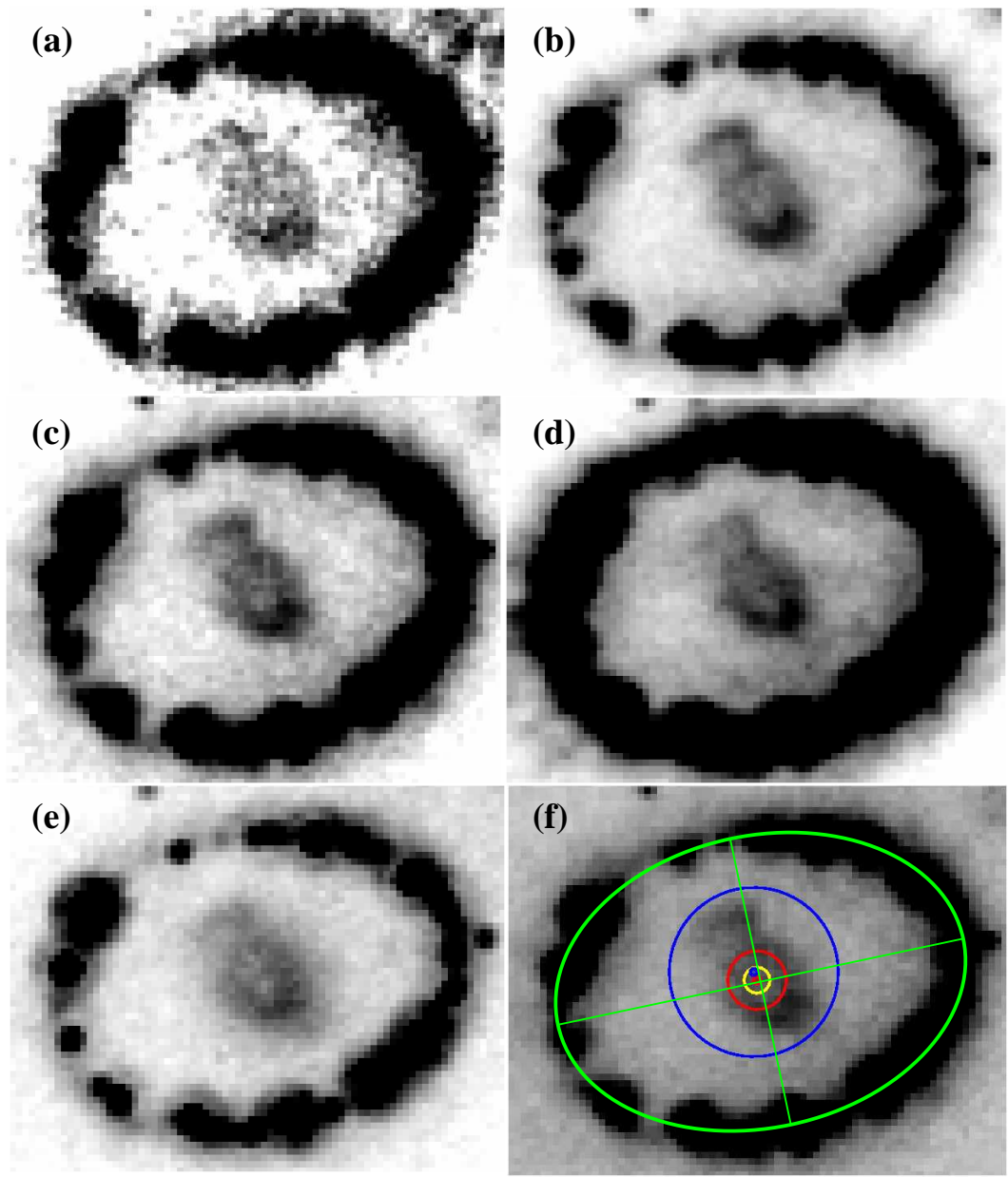}
\caption{{\it ACS/HRC} images of SN 1987A in five filters:  (a) {\it HRC}
UV - F330W, (b) Johnson B - F435W, (c) Johnson V - F555W, (d) SDSS r -
F625W, and (e) Broad I - F814W.  No central point source is detected
in any band. Part (f) shows the F555W image with three determinations
of the center of the debris: by eye (red dot), by fitting a circle to
the central debris (blue), and by fitting an ellipse to the inner ring
of the SNR (green).  The yellow circle has a radius of two pixels and
shows the uncertainty in determining the centerpoint.  The red circle
has a radius of 4.5 pixels and encloses the maximum distance that a
compact object may have travelled from the center, with a kick
velocity $\le 1000$ km/s.} \label{images}
\end{figure}

\clearpage

\begin{figure}
\epsscale{1}
\plotone{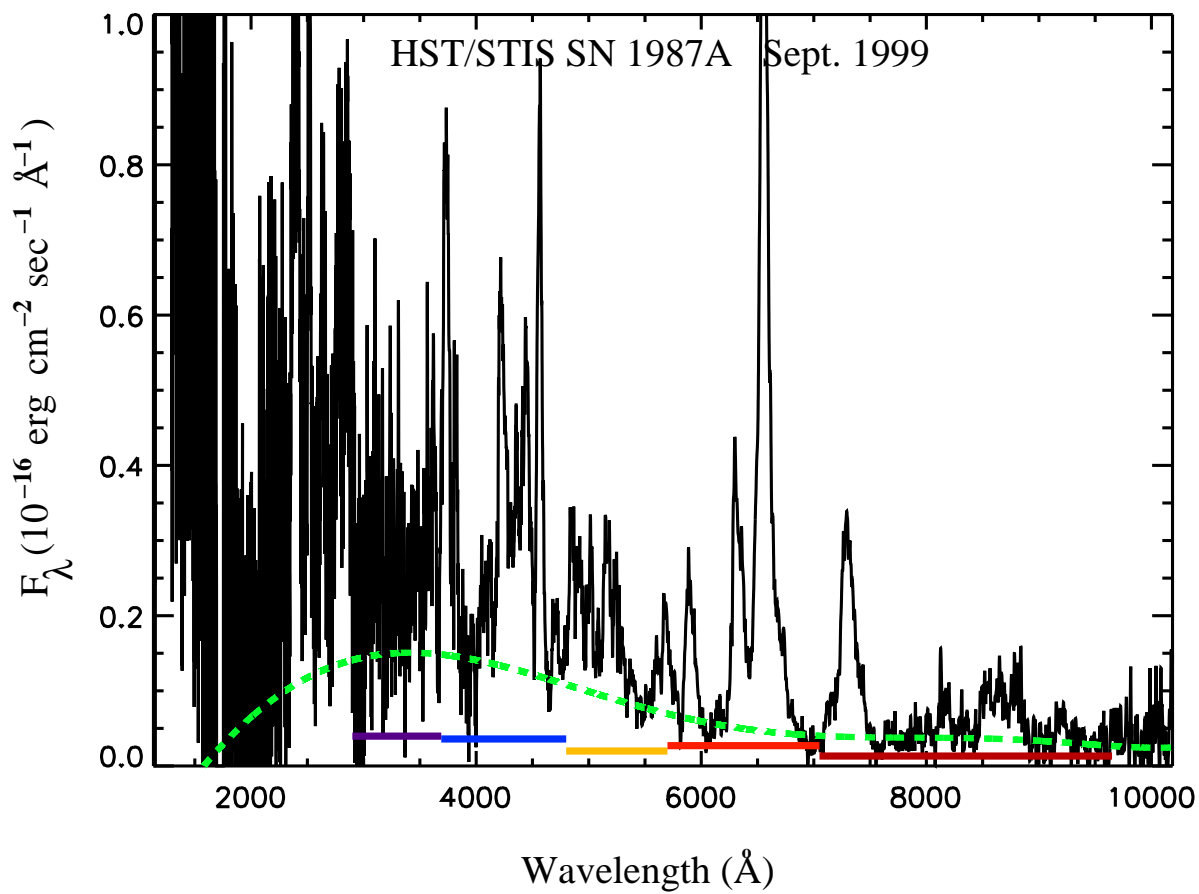}
\caption{The 1999 September spectrum of SN 1987A, corrected for
interstellar extinction.  The fit to a possible underlying continuum 
is shown as the green dashed line.  Also shown are the 2003
November {\it
ACS/HRC} limits, plotted as horizontal colored bars that span the applied filter
width.  From left to right, these are the F330W, F435W, F555W, F625W,
and F814W filter bands.  The imaging data place a more stringent limit on a
continuum point source in the remnant than do the spectral data.}
\label{final_cont_image}
\end{figure} 

\clearpage

\begin{figure}
\epsscale{1}
\plotone{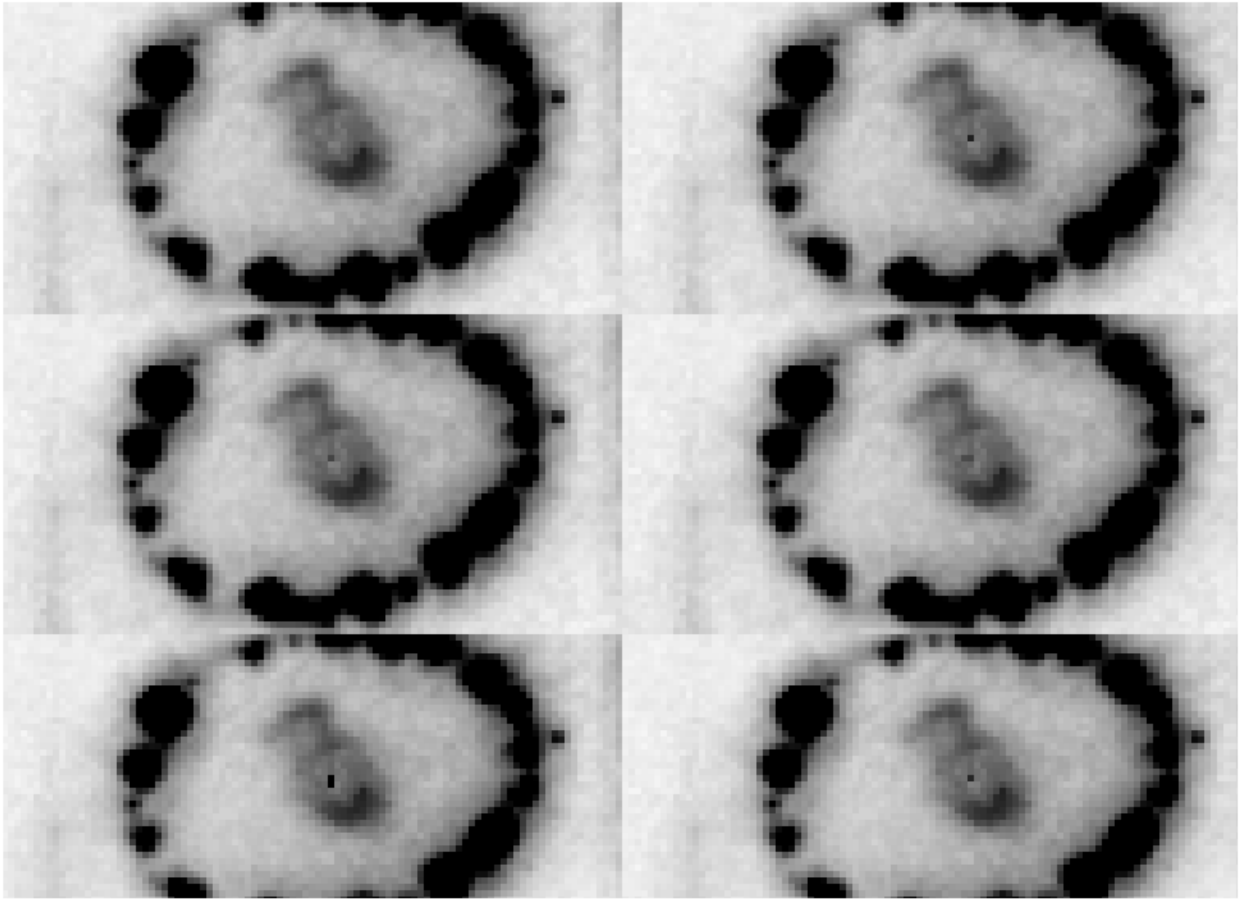}
\caption{{\it ACS/HRC} image of SN 1987A in the Johnson B filter
(F435W) with inserted point sources.  The inserted sources increase in
total counts from left to right and top to bottom.  The top left
inserted source is not easily detectable, and is therefore the upper limit in
the B-band.}\label{f435_fakes}
\end{figure}

\clearpage  

\begin{figure}
\epsscale{1}
\plotone{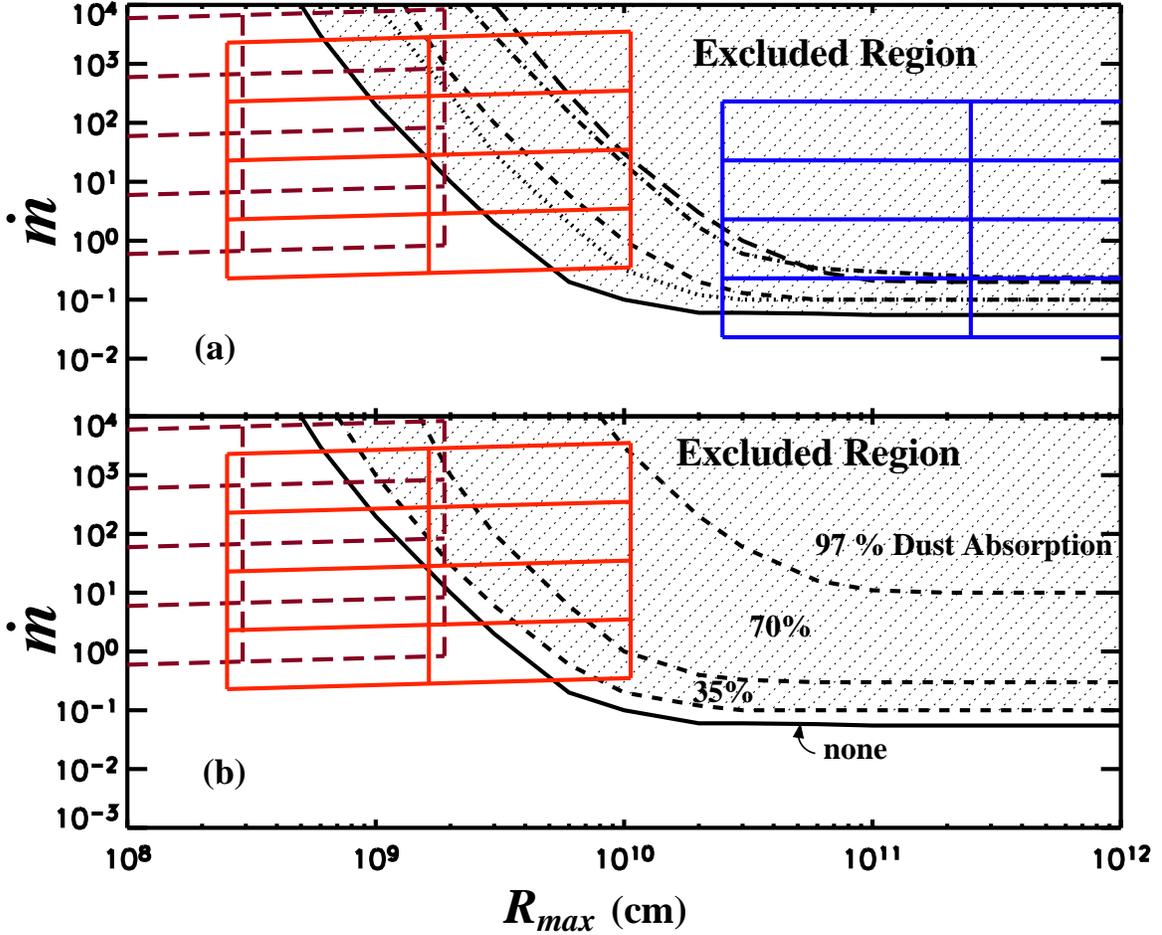}
\caption{Constraints on thin disk accretion rate and disk radius given by our
observed upper limit.  (a) The solid line shows
the limit imposed by the UV observations.  The shaded region shows disk radii and accretion rates that are excluded by the
observed limit on counts in the UV.  Limits from other bands are also shown:
the dotted, short-dashed, dot-dashed, and long-dashed lines correspond
to the limits from the B-band, V-band, R-band, and I-band, respectively.
The blue grid
shows accretion disks predicted by the PHN fallback
model.  The red grids show the accretion disks predicted by the
MPH fallback model; the solid grid shows the results for
hot disks ($T=10^6$ K), while the darker dashed grid shows the results
for cooler disks ($T=10^4$ K).  In each grid, verticle lines
correspond to initial disk radii (from left to right) of $10^6$ cm,
$10^7$ cm, and $10^8$ cm.  Horizontal lines correspond to intial
disk masses (from top to bottom) of $10^{-1} M_{\odot}$, $10^{-2}
M_{\odot}$, $10^{-3} M_{\odot}$, $10^{-4} M_{\odot}$, and $10^{-5}
M_{\odot}$. (b) The effect of dust absorption on the upper limit.  As
in (a), the solid curve is the UV-band limit.  The dashed lines show the
UV-band limit in the case of 35\%, 70\%, and 97\% dust absorption
within the remnant.  The red grids show the accretion disks predicted
by the MPH fallback model, as in (a).} \label{disks_uvcolor}
\end{figure} 


\clearpage

\begin{thebibliography}{}
\bibitem[Arnett et al.(1989)]{arn89} Arnett, W. D., Bahcall, J. N., Kirshner, R. P., 
	{\&} Woosley, S. E. 1989, \araa, 27, 629
\bibitem[Bath {\&} Pringle(1981)]{bat81} Bath, G. T., {\&} Pringle, J. E. 
	1981, \mnras, 194, 967
\bibitem[Blandford, Applegate, \& Hernquist(1983)]{bla83} Blandford,
R. D., Applegate, J. H., \& Hernquist, L. 1983, \mnras, 204, 1025
\bibitem[Blondin(1986)]{blo86} Blondin, J. M. 1986, \apj, 308, 755
\bibitem[Borkowski et al.(1997)]{bor97} Borkowski, K. J., de Kool, M., McCray, R., {\&} Wooden, 
	D. H. 1997, Amer. Astron. Soc. Meet. 191, {\#}40.12
\bibitem[Bouchet et al.(2003)]{bou03} Bouchet, P., De Buizer, J. M.,
  Suntzeff, N. B., Danziger, I. J., Hayward, T. L., Telesco, C. M., \&
  Packham, C. 2003, \apj, 611, 394
\bibitem[Bouchet {\&} Danziger(1993)]{bou93} Bouchet, P., {\&} Danziger, I. J. 1993, \aap, 273, 451
\bibitem[Bouchet et al.(1996)]{bou96} Bouchet, P., Danziger, I. J., Gouiffes, C., della Valle, M., 
	{\&} Monetti, A. 1996, in Supernovae and Supernova Remnants, IAU Colloq. 145, ed. R.
	McCray {\&} Z. Wang (Cambridge: Cambridge Univ. Press), 201
\bibitem[Brown {\&} Bethe(1994)]{brb94} Brown, G. E., {\&} Bethe, H. A. 
	1994, \apj, 423, 659
\bibitem[Brown {\&} Weingartner(1994)]{bro94} Brown, G. E., {\&} 
	Weingartner, J. C. 1994, \apj, 436, 843
\bibitem[Cannizzo, Lee, {\&} Goodman(1990)]{can90} Cannizzo, J. K., Lee, H. M.,
	{\&} Goodman, J. 1990, \apj, 351, 38
\bibitem[Chakrabarty et al.(2001)]{cha01} Chakrabarty, D., Pivovaroff,
  M. J., Hernquist, L. E., Heyl, J. S., \& Narayan, R. 2001, \apj,
  548, 800
\bibitem[Chatterjee, Hernquist, {\&} Narayan(2000)]{chat00} Chatterjee, P., 
	Hernquist, L. E., {\&} Narayan, R. 2000, \apj, 534, 373
\bibitem[Chevalier(1989)]{che89} Chevalier, R. A. 1989, \apj, 346, 847
\bibitem[Chevalier {\&} Kirshner(1978)]{che78} Chevalier, R. A., {\&} Kirshner, 
	R. P. 1978, \apj, 219, 931
\bibitem[Chugai et al.(1997)]{chu97} Chugai, N. N., Chevalier, R. A., 
	Kirshner, R. P., {\&} Challis, P. M. 1997, \apj, 483, 925
\bibitem[Collins et al.(1999)]{col99} Collins, T. J. B., Frank, A., Bjorkman, J. E., {\&}
	Livio, M. 1999, \apj, 512, 322
\bibitem[Colpi, Shapiro, {\&} Wasserman(1996)]{col96} Colpi, M., Shapiro, S. 
	L., {\&} Wasserman, I. 1996, \apj, 470, 1075
\bibitem[Corbel et al.(1999)]{cor99} Corbel, S., Chapuis, C., Dame,
  T. M., \& Durouchoux, P. 1999, \apj, 526, L29
\bibitem[Crotts \& Heathcote(1991)]{cro91} Crotts, A. P., \& Heathcote, S. R. 1991, \nat, 
	350, 683
\bibitem[Dolphin(2000)]{dol00} Dolphin, A. E. 2000, \pasp, 112, 1397
\bibitem[Eardley et al.(1978)]{ear78} Eardley, D. M., Lightman, A. P., 
	Payne, D. G., {\&} Shapiro, S. L. 1978, \apj, 224, 53
\bibitem[Fabian {\&} Rees(1988)]{fab88} Fabian, A. C., {\&} Rees, M. J. 1988, \nat, 335, 50
\bibitem[Fesen et al.(2002)]{fes02} Fesen, R. A., Chevalier, R. A., Holt, S. S., \& Tananbaum, 
	H. 2002, in Neutron Stars in Supernova Remnants, ASP Conf. Series Vo. 271, 
	ed. P. O. Slane \& B. M. Gaensler (San Francisco: ASP), 305
\bibitem[Fischera, Tuffs, {\&} V{\"{o}}lk(2002a)]{fis02a} Fischera, J., Tuffs, R. J., {\&} V\"{o}lk, H. J. 2002a, \aap, 386, 517
\bibitem[Fischera, Tuffs, {\&} V{\"{o}}lk(2002b)]{fis02b} Fischera, J., Tuffs, R. J., {\&} V\"{o}lk, H. J. 2002b, \aap, 395, 189
\bibitem[Fitzpatrick(1986)]{fit86} Fitzpatrick, E. L. 1986, \aj, 92, 1068
\bibitem[Frank, King, {\&} Raine(1992)]{fra92} Frank, J., King, A. R., {\&} 
	Raine, D. J. 1992, Accretion Power in Astrophysics (Cambridge: 
	Cambridge Univ. Press)
\bibitem[Fransson \& Kozma(2002)]{fra02} Fransson, C., \& Kozma, C. 2002, New Astronomy Reviews, 46, 487
\bibitem[Fruchter \& Hook(2002)]{fru02} Fruchter, A. S., \& Hook,
  R. N. 2002, \pasp, 114, 144
\bibitem[Fryer(1999)]{fryer99} Fryer, C. L. 1999, \apj, 522, 413
\bibitem[Fryer, Colgate, {\&} Pinto(1999)]{fry99} Fryer, C. L., Colgate, S. 
	A., {\&} Pinto, P. A. 1999, \apj, 511, 885
\bibitem[Gaensler, Gotthelf, \& Vasisht(1999)]{gae99} Gaensler, B. M.,
Gotthelf, E. V., \& Vasisht, G. 1999, \apj, 526, L37
\bibitem[Gaensler et al.(1997)]{gae97} Gaensler, B. M., Manchester, R. N., 
	Staveley-Smith, L., Tzioumis, A. K., Reynolds, J. E., {\&} Kesteven, 
	M. J. 1997, \apj, 479, 845
\bibitem[Gotthelf, Petre, \& Hwang(1997)]{got97} Gotthelf, E. V.,
  Petre, R., \& Hwang, U. 1997, \apj, 487, L175
\bibitem[Gotthelf \& Vasisht(1997)]{gotthelf97} Gotthelf, E. V., \&
Vasisht, G. 1997, \apj, 486, L133
\bibitem[Gotthelf et al.(2000)]{got00} Gotthelf, E. V., Vasisht, G.,
  Boylan-Kolchin, M., \& Torii, K. 2000, \apj, 542, L37
\bibitem[Gotthelf, Vasisht, \& Dotani(1999)]{got99} Gotthelf, E. V.,
  Vasisht, G., \& Dotani, T. 1999, \apj, 522, 49
\bibitem[Gould(1994)]{gou94} Gould, A. 1994, \apj, 425, 51
\bibitem[Hameury et al.(1998)]{ham98} Hameury, J.-M., Menou, K.,
	Dubus, G., Lasota, J.-P., {\&} Hur\'{e}, J.-M. 1998, \mnras, 298, 1048
\bibitem[Harnden \& Seward(1984)]{har84} Harnden, F. R., \& Seward,
  F. D. 1984, \apj, 283, 279
\bibitem[Harrison, Lyne, {\&} Anderson(1993)]{har93} Harrison, P. A., Lyne, 
	A. G., {\&} Anderson, B. 1993, \mnras, 261, 113
\bibitem[Heger, Langer, {\&} Woosley(2000)]{heg00} Heger, A., Langer,
	N., {\&} Woosley, S. E. 2000, \apj, 528, 368
\bibitem[Heger, Woosley, \& Spruit(2004)]{heg04} Heger, A., Woosley,
S. E., \& Spruit, H. C. 2004, astro-ph/0409422
\bibitem[Holtzman et al.(1995)]{hol95} Holtzman, J. A., Burrows, C. J., 
	Casertano, S., Hester, J. J., Trauger, J. T., Watson, A. M., {\&} Worthey, 
	G. 1995, \pasp, 107, 1065
\bibitem[Houck {\&} Chevalier(1991)]{hou91} Houck, J. C., {\&} Chevalier, 
	R. A. 1991, \apj, 376, 234
\bibitem[Howarth(1983)]{how83} Howarth, I. D. 1983, \mnras, 203, 301
\bibitem[Hurley et al.(2000)]{hur00} Hurley, K., et al. 2000, \apj,
  528, L21
\bibitem[Illarionov {\&} Sunyaev(1975)]{ill75} Illarionov, A. F., {\&} Sunyaev, 
	R. A. 1975, \aap, 39, 185
\bibitem[Jaroszy\'{n}ski, Abramowicz, \& Paczy\'{n}ski(1980)]{jar80}
Jaroszy\'{n}ski, M., Abramowicz, M. A., \& Paczy\'{n}ski, B. 1980,
Acta Astronomica, 30, 1
\bibitem[Joss et al.(1988)]{jos88} Joss, P. C., Podsiadlowski, P., Hsu, J. J. L., {\&} Rappaport, 
	S. 1988, \nat, 331, 237
\bibitem[Kaart et al.(2001)]{kaa01} Kaart, P., et al. 2001, \apj, 321, L29
\bibitem[Kirshner et al.(1987)]{kir87} Kirshner, R. P., Sonneborn, G., 
	Crenshaw, D. M., {\&} Nassiopoulos, G. E. 1987, \apj, 320, 602
\bibitem[Koptsevich et al.(2001)]{kop01} Koptsevich, A. B., Pavlov,
  G. G., Zharikov, S. V., Sokolov, V. V., Shibanov, Y. A., \& Kurt,
  V. G. 2001, \aap, 370, 1004
\bibitem[Kozma {\&} Fransson(1998)]{koz98} Kozma, C., {\&} Fransson, C. 1998, \apj, 496, 946
\bibitem[Kristian(1991)]{kri91} Kristian, J. 1991, \nat, 349, 747
\bibitem[Kristian et al.(1989)]{kri89} Kristian, J., et al. 1989, \nat, 338, 234
\bibitem[de Loore {\&} Vanbeveren(1992)]{loo92} de Loore, C., {\&} Vanbeveren, D. 1992, \aap, 260, 273
\bibitem[Lucy et al.(1989)]{luc89} Lucy, L. B., Danziger, I. J., Gouiffes, 
	C., {\&} Bouchet, P. 1989, in Structure {\&} Dynamics of the Interstellar 
	Medium, IAU Colloq. 120, ed. G. Tenorio-Tagle, M. Mole, {\&} J. 
	Melnick (Berlin: Springer), 164
\bibitem[Lucy et al.(1991)]{luc91} Lucy, L. B., Danziger, I. J., Gouiffes, 
	C., {\&} Bouchet, P. 1991, in Supernovae, ed. S. E. Woosley (Berlin: Springer), 82
\bibitem[Lundqvist {\&} Fransson(1996)]{lun96} Lundqvist, P., {\&} Fransson, C. 1996, \apj, 464, 924
\bibitem[Lundqvist et al. (2001)]{lun01} Lundqvist, P., Kozma, C., Sollerman, J., {\&} Fransson, C. 2001, \aap, 374, 629
\bibitem[Lundqvist et al. (1999)]{lun99} Lundqvist, P., Sollerman, J., Kozma, C., Larsson, B., Spyromilio, J., 
        Crotts, A. P. S., Danziger, J., {\&} Kunze, D. 1999, \aap, 347, 500
\bibitem[Mack et al.(2003)]{mac03} Mack, J. et al. 2003, ``ACS Data
	Handbook,'' Version 2.0, (Baltimore: STScI)
\bibitem[Marsden et al.(1997)]{mar97} Marsden, D., et al. 1997, \apj,
  491, L39
\bibitem[Marshall et al.(1998)]{mar98} Marshall, F. E., Gotthelf,
E. V., Zhang, W., Middleditch, J., \& Wang, Q. D. 1998, \apj, 499, L179
\bibitem[McCray(1993)]{mcc93} McCray, R. 1993, \araa, 31, 175
\bibitem[Menou et al.(1999)]{men99} Menou, K., Esin, A. A., Narayan, R., 
	Garcia, M. R., Lasota, J.-P., {\&} McClintock, J. E. 1999, \apj, 520, 276
\bibitem[Menou, Perna, {\&} Hernquist(2001)]{men01} Menou, K., Perna, R., {\&}
	Hernquist, L. 2001, \apj, 559, 1032
\bibitem[Mereghetti et al.(2002)]{mer02} Mereghetti, S., Bandiera, R.,
Bocchino, F., \& Israel, G. L. 2002, \apj, 574, 873
\bibitem[Mereghetti, Bignami, \& Caraveo(1996)]{mer96} Mereghetti, S.,
  Bignami, G. F., \& Caraveo, P. A. 1996, \apj, 464, 842
\bibitem[Middleditch, Pennypacker, \& Burns(1987)]{mid87} Middleditch,
  J., Pennypacker, C. R., \& Burns, M. S. 1987, \apj, 315, 142
\bibitem[Middleditch et al.(2000)]{mid00} Middleditch, J., et al. 2000, New Astronomy, 5, 243
\bibitem[Mignani et al.(2000)]{mig00} Mignani, R. P., Pulone, L.,
  Marconi, G., Iannicola, G., \& Caraveo, P. A. 2000, \aap, 355, 603
\bibitem[Mitchell et al.(2002)]{mit02} Mitchell, R. C., Baron, E., Branch, D., Hauschildt, P. H., Nugent, P. 
	E., Lundqvist, P., Blinnikov, S., {\&} Pun, C. S. J. 2002, \apj, 574, 293
\bibitem[Narayan, Mahadevan, {\&} Quataert(1998)]{nar98} Narayan, R., Mahadevan,
	 R., {\&} Quataert, E. 1998, Theory of Black Hole Accretion Discs, ed.
	 M. Abramowicz, G. Bjornsson, {\&} J. Pringle (Cambridge: Cambridge Univ. 
	Press)
\bibitem[Natta {\&} Panagia(1984)]{nat84} Natta, A., {\&} Panagia, N. 1984, \apj, 287, 228
\bibitem[Nomoto, Shigeyama, {\&} Hashimoto(1987)]{nom87} Nomoto, K., Shigeyama, T., 
        \& Hashimoto, M. 1987, in SN 1987A, ed. I. J. Danziger (ESO: Garching) p. 325
\bibitem[\"{O}gelman {\&} Alpar(2004)]{oge04} \"{O}gelman, H., {\&}
	Alpar, M. A. 2004, \apj, 603, L33
\bibitem[Panagia(1999)]{pan99} Panagia, N. 1999, in New Views of the Magellanic Cloud, IAU Symp. 
	190, ed. Y.-H. Chu, N. Suntzeff, J. Hesser, {\&} D. Bohlender (San Francisco: ASP), 549
\bibitem[Panagia et al.(1991)]{pan91} Panagia, N., Gilmozzi, R., Macchetto, F., Adorf, 
        H.-M., {\&} Kirshner, R. P. 1991, \apj, 380, L23
\bibitem[Panagia et al.(2000)]{pan00} Panagia, N., Romaniello, M., Scuderi, S., {\&} Kirshner, R. P. 
	2000, \apj, 539, 197
\bibitem[Park et al.(2004)]{par04} Park, S., Zhekov, S. A., Burrows,
D. N., Michael, E., McCray, R., Garmire, G. P., {\&} Hasinger,
G. 2004, Advances in Space Research, 33, 386
\bibitem[Pavlov, Stringfellow, \& Cordova(1996)]{pav96} Pavlov, G. G.,
Stringfellow, G. S., \& Cordova, F. A. 1996, \apj, 467, 370
\bibitem[Pavlov et al.(2000)]{pav00} Pavlov, G. G., Zavlin, V. E., Aschenbach, B., Tr\"{u}mper, J., 
        \& Sanwal, D. 2000, \apj, 531, L53
\bibitem[Pavlov, Zavlin, \& Sanwal(2002)]{pav02} Pavlov, G. G.,
  Zavlin, V. E., \& Sanwal, D. 2002, in Neutron Stars, Pulsars and
  Supernova Remnants, Proc. of the 270-th Heraeus Seminar,
  ed. W. Becker, H. Lesch \& J. Tr\"{u}mper (MPE Report 278), 283 (astro-ph/0206024)
\bibitem[Pavlov et al.(2002)]{pavlov02} Pavlov, G. G., Zavlin, V. E.,
Sanwal, D., \& Tr\"{u}mper, J. 2002, \apj, 569, L95
\bibitem[Perna, Hernquist, {\&} Narayan(2000)]{per00} Perna, R., Hernquist, L. 
	E., {\&} Narayan, R. 2000, \apj, 541, 344
\bibitem[Petre, Becker, \& Winkler(1996)]{pet96} Petre, R., Becker,
  C. M., \& Winkler, P. F. 1996, \apj, 541, 344
\bibitem[Podsiadlowski(1989)]{pod89} Podsiadlowski, P. 1989, Ph.D. Thesis, Massachusetts 
	Institute of Technology
\bibitem[Pringle(1974)]{pri74} Pringle, J. E. 1974, Ph.D. Thesis,
	University of Cambridge
\bibitem[Rho \& Petre(1997)]{rho97} Rho, J., \& Petre, R. 1997, \apj,
  484, 828
\bibitem[Sanduleak(1969)]{san69} Sanduleak, N. 1969, Contr. CTIO, 89, 1
\bibitem[Sasaki et al.(2004)]{sas04} Sasaki, M., Plucinsky, P. P.,
Gaetz, T. J., Smith, R. K., Edgar, R. J., \& Slane, P. O. 2004, \apj,
617, 322
\bibitem[Savage {\&} Mathis(1979)]{sav79} Savage, B. D., {\&} Mathis,
J. S. 1979, \araa, 17, 73
\bibitem[Scuderi et al.(1996)]{scu96} Scuderi S., Panagia, N.,
Gilmozzi, R., Challis, P. M., {\&} Kirshner, R. P. 1996, \apj, 465, 956
\bibitem[Serafimovich et al.(2004)]{ser04} Serafimovich, N. I.,
  Shibanov, Yu. A., Lundqvist, P., \& Sollerman, J. 2004, \aap, 425, 1041
\bibitem[Seward \& Harnden(1994)]{sew94} Seward, F. D., \& Harnden,
  F. R. J. 1994, \apj, 421, 581
\bibitem[Seward, Harnden, \& Helfand(1984)]{sew84} Seward, F. D.,
Harnden, F. R., \& Helfand, D. J. 1984, \apj, 287, L19
\bibitem[Shakura {\&} Sunyaev(1973)]{sha73} Shakura, N. I., {\&} Sunyaev, R. A.
	 1973, \aap, 24, 337
\bibitem[Shapiro, Lightman, \& Eardley(1976)]{sha76} Shapiro, S. L.,
  Lightman, A. P., \& Eardley, D. M. 1976, \apj, 204, 187
\bibitem[Shibanov et al.(2003)]{shi03} Shibanov, Y. A., Koptsevich,
  A. B., Sollerman, J., \& Lundqvist, P. 2003, \aap, 406, 645
\bibitem[Shklovskii (1979)]{shk79} Shklovskii, I. S. 1979, \nat, 279, 703
\bibitem[Shtykovskiy et al.(2004)]{sht04} Shtykovskiy, P., Lutovinov,
A., Gilfanov, M., \& Sunyaev, R. 2004, preprint (astro-ph/0411731)
\bibitem[Soderberg, Challis, {\&} Suntzeff(1999)]{sod99} Soderberg,
A. M., Challis, P. M., {\&} Suntzeff, N. B. 1999, BAAS, 31, 977
\bibitem[Sollerman et al.(2000)]{sol00} Sollerman, J., et al. 2000,
  \apj, 537, 861
\bibitem[Sonneborn, Altner, {\&} Kirshner(1987)]{son87} Sonneborn, G., Altner, 
	B., {\&} Kirshner, R. P. 1987, \apj, 323, L35
\bibitem[Stetson(1987)]{ste87} Stetson, P. 1987, \pasp, 99, 191
\bibitem[Suntzeff {\&} Bouchet(1990)]{sun90} Suntzeff, N. B., {\&}
Bouchet, P. 1990, \aj, 99, 650
\bibitem[Suntzeff et al.(1992)]{sun92} Suntzeff, N. B., Phillips, M. M., Elias, J. H., DePoy, D. L., {\&}
        Walker, A. R. 1992, \apj, 384, L33
\bibitem[Tananbaum(1999)]{tan99} Tananbaum, H. 1999, IAU Circ., No. 7246
\bibitem[Thorsett(1992)]{tho92} Thorsett, S. E. 1992, \nat, 356, 690
\bibitem[Torii et al.(1998)]{torii98} Torii, K. I., Kinugasa, K.,
Toneri, T., Asanuma, T., Tsunemi, H., Dotani, T., Mitsuda, K.,
Gotthelf, E. V., \& Petre, R. 1998, \apj, 494, L207
\bibitem[Torii et al.(1998)]{tor98} Torii, K. I., Kinugasa, K.,
  Katayama, K., Tsunemi, H., \& Yamauchi, S. 1998, \apj, 503, 843
\bibitem[Tuohy \& Garmire(1980)]{tuo80} Tuohy, I., \& Garmire,
  G. 1980, \apj, 239, 107
\bibitem[van den Bergh \& Kamper(1983)]{van83} van den Bergh, S., \&
Kamper, K. W. 1983, \apj, 268, 129
\bibitem[Vasisht et al.(1994)]{vas94} Vasisht, G., Kulkarni, S. R.,
Frail, D. A., \& Greiner, J. 1994, \apj, 431, L35
\bibitem[Walborn et al.(1987)]{wal87} Walborn, N. R., Lasker, B. M., Laidler, V. G., 
	\& Chu, Y.-H. 1987, \apj, 321, L41
\bibitem[Wang et al.(1996)]{wan96} Wang, L., et al. 1996, \apj, 466, 998
\bibitem[Wang et al.(2002)]{wan02} Wang, L., et al. 2002, \apj, 579, 671
\bibitem[Wang \& Chakrabarty(2002)]{wanc02} Wang, Z.-X., \&
  Chakrabarty, D. 2002, in Neutron Stars in Supernova Remnants, ASP Conf. Series Vo. 271, 
	ed. P. O. Slane \& B. M. Gaensler (San Francisco: ASP), 297 (astro-ph/0112125)
\bibitem[Wang \& Gotthelf(1998)]{wan98} Wang, Q. D., \& Gotthelf,
  E. V. 1998, \apj, 509, L109
\bibitem[West(1987)]{wes87} West, R. M. 1987, \aap, 177, L1
\bibitem[White \& Malin(1987)]{whi87} White, G. L., \& Malin,
  D. F. 1987, \nat, 327, 36
\bibitem[Winkler \& Kirshner(1985)]{win85} Winkler, P. F., \&
Kirshner, R. P. 1985, \apj, 299, 981
\bibitem[Winkler, Kirshner, \& Irwin(1986)]{win86} Winkler, P. F.,
Kirshner, R. P., \& Irwin, M. M. 1986, \baas, 18, 1053 
\bibitem[Woods et al.(1999)]{wood99} Woods, P. M., Kouveliotou, C.,
  van Paradijs, J., Finger, M. H., \& Thompson, C. 1999, \apj, 518, L103
\bibitem[Woosley(1988)]{woo88} Woosley, S. E. 1998, \apj, 330, 218
\bibitem[Zampieri et al.(1998)]{zam98} Zampieri, L., Colpi, M., Shapiro, S. 
	L., {\&} Wasserman, I. 1998, \apj, 505, 876
\bibitem[Zavlin \& Pavlov(2004a)]{zav04a} Zavlin, V. E., \& Pavlov,
  G. G. 2004, in Proc. of the {\it XMM}-Newton EPIC Consortium,
  Mem.S.A.It., 75, 458 (astro-ph/0312326)
\bibitem[Zavlin \& Pavlov(2004b)]{zav04b} Zavlin, V. E., \& Pavlov,
  G. G. 2004, \apj, 616, 452
\end{thebibliography}
\end{document}